\documentclass[%
 reprint,
 amsmath,amssymb,
 aps,
 showkeys
]{revtex4-2}
\usepackage{gensymb}
\usepackage{booktabs}
\usepackage{graphicx}
\graphicspath{{./Figures/}}
\usepackage{dcolumn}
\usepackage{multirow}
\usepackage{bm}
\usepackage{times}

\usepackage[colorlinks = true,
            linkcolor = blue,
            urlcolor  = blue,
            citecolor = blue,
            anchorcolor = blue]{hyperref}

\newcommand{\adisl}{$\langle 110 \rangle/2$ }
\newcommand{\bdisl}{$\langle 112 \rangle/2$ }
\newcommand{\andisl}{$\langle 110 \rangle/2$}
\newcommand{\bndisl}{$\langle 112 \rangle/2$}

\newcommand{\beginsupplement}{%
        \setcounter{table}{0}
        \renewcommand{\thetable}{S\arabic{table}}%
        \setcounter{figure}{0}
        \renewcommand{\thefigure}{S\arabic{figure}}%
        \setcounter{section}{0}
        \renewcommand{\thesection}{\Roman{section}}%
     }

\begin{document}

\preprint{APS/123-QED}

\title{A ``Magnetic'' Machine Learning Interatomic Potential for  Nickel}

\author{Xiaoguo Gong}
\author{Zhuoyuan Li}
\author{A. S. L. Subrahmanyam Pattamatta}%
\author{Tongqi Wen}\email{tongqwen@hku.hk}
\author{David J. Srolovitz}\email{srol@hku.hk}

\affiliation{Department of Mechanical Engineering, The University of Hong Kong, Hong Kong Special Administrative Region of China}


\begin{abstract}

  Nickel (Ni) is a magnetic transition metal with two allotropic phases, stable face-centered cubic (FCC) and metastable hexagonal close-packed (HCP), widely used in structural applications. 
  Magnetism affects many mechanical and defect properties, but spin-polarized density functional theory (DFT) calculations are computationally inefficient for studying material behavior requiring large system sizes and/or long simulation times. Here we develop a ``magnetism-hidden'' machine-learning Deep Potential (DP) model for   Ni without a descriptor for magnetic moments, using training datasets derived from spin-polarized DFT calculations. 
  The ``magnetism-hidden'' DP-Ni model exhibits high transferability and representability for a wide-range of FCC and HCP properties, including (finite-temperature) lattice parameters, elastic constants, phonon spectra, and many defects. 
  As an example of its applicability, we investigate the Ni FCC-HCP allotropic phase transition under (high-stress) uniaxial tensile loading. 
  The DP model for magnetic Ni facilitates accurate large-scale atomistic simulations for complex mechanical behavior and serves as a foundation for developing interatomic potentials for Ni-based superalloys and other multi-principal component alloys.

\end{abstract}

\maketitle


\section{Introduction}
\label{sec:intro}

Although most applications of nickel (a soft ferromagnet below 627 K) do not focus on their magnetic properties, many of its non-magnetic properties depend sensitively on its electronic spin degrees of freedom. 
Inclusion of magnetic degrees of freedom impacts the $ab$ $initio$ prediction of phase stability~\cite{PhysRevB.67.035116,Zelen_Prb}, vacancy and self-interstitial formation energies~\cite{HARGATHER201417,PhysRevB.74.064111,MIZUNO2001386,GongYl_Prb}, elastic moduli~\cite{PhysRevB.67.035116,guo2000gradient}, stacking fault energies~\cite{Chandran_Jap, KUMAR2018424, Zhang_Prb,Brandl_Prb}, and mechanical properties.
For example, Ni elastic constants, calculated via density-functional theory (DFT) without spin polarization lead to errors of $\sim$23\%~\cite{guo2000gradient} compared with the experiment. 
Similarly, stacking fault energies calculated with and without spin degrees of freedom differ by  24-50\%~\cite{Chandran_Jap,KUMAR2018424}. 
Hence, magnetism is important for a wide range of non-magnetic (structural) properties. 
Here, we propose an approach to developing an interatomic potential for the structural and thermodynamic properties of magnetic metals and their crystal defects and implement it for the industrially important metal, nickel. 
We do this with a machine learning (ML) potential framework which is trained against DFT calculations that include magnetic degrees of freedom.

While DFT provides a highly accurate, quantum mechanics-based approach to understanding the structure and properties of Ni, its applicability to the properties of defects and finite temperature behavior is limited by the large computational demands required for reasonable system size and time scales. 
DFT calculations involving $10^{3-5}$  spin-polarized atoms and time scale $>1$ nanosecond are heroic. 
Empirical or semi-empirical interatomic potentials are routinely employed to enable the simulation of the properties of metals and their defects on these scales.
Over the past fifty years, dozens of Ni potentials have been developed (e.g., see~\cite{IPR,KIM}) and achieved some successes in explaining experimental observations and predicting material behavior. 
However, the transferability and accuracy of these potentials are limited by their fixed functional form;  this concern is particularly acute in determining the properties of non-equilibrium structures, such as, hexagonal close-packed (HCP) Ni. 
We benchmarked the basic properties of face-centered cubic (FCC) and HCP Ni using various interatomic potentials, including 8 embedded-atom method (EAM) and 10 modified embedded-atom method (MEAM) potentials (see Supplementary Table S1). 
All potentials display significant discrepancies in simple properties, such as the elastic constants ($C_{ij}$) of metastable HCP Ni (possibly due to the limited availability of fitting data);  these deviations can be as high as 41\% (see  $C_{13}$, $C_{33}$,  $C_{44}$  in Table S1).
This makes accurate prediction of the mechanical (elastic and plastic) deformation of HCP Ni (as well as other non-FCC phases) and phase transitions in Ni using these potentials unreliable.
Indeed, allotropic phases significantly impact the strength/toughness of many metals~\cite{pattamatta2022allotropy,krygier2019extreme,luo2019plastic,YU20179}. 
A recent example of such transformations was seen in heavily deformed, nanocrystalline (grain refined) Ni~\cite{luo2019plastic, guo2021plastic}; at a grain size of $\sim$17 nm, 5-10\% of grains transform to  HCP (HCP Ni is harder and stronger than  FCC Ni)~\cite{luo2019plastic}.
HCP Ni is also widely observed in thin hetero-epitaxial  films~\cite{tian2005hexagonal,higuchi2011preparation}. 
HCP nickel formation thermodynamics and its transformation behavior are unclear~\cite{pattamatta2022allotropy,tian2005hexagonal,luo2019plastic,guo2021plastic}.

One approach to achieving accurate, efficient predictions is through atomistic simulations employing machine-learning based interatomic potentials (see ~\cite{wen2021specialising} for a recent example for titanium). 
Application of a similar ML approach for Ni~\cite{zuo2020performance,li2018quantum}, that does not include spin polarization, has proven unreliable for several important properties (see Section~\ref{sec:Results and Discussion}).
One strategy for developing ML potentials that capture the influence of magnetism moments is the incorporation of explicit descriptors of the magnetic degrees of freedom within the ML potential. 
For example,  incorporating magnetic moments in the ML potentials (e.g.,  Fe and Mn-containing systems) greatly improves the accuracy of the prediction of thermodynamic and structural properties of magnetic metals~\cite{suzuki2023high,novikov2022magnetic,yu2022time,zhang2023machine,egorov2023magnetic,toda2022generalized,chapman2022machine,yu2022spin}. 
However, this implementation necessitates, larger, more costly, DFT training datasets, and increases the complexity of ML potential training and generates ML potentials that are more computationally costly to use. 

Here, we develop ML potentials that incorporate the effects of magnetism without an explicit description of the magnetic degrees of freedom; such potentials are applicable to the robust prediction of non-magnetic properties.
Specifically, we develop an ML deep potential (DP)~\cite{zhang2020dp,wen_2022_mf}  for the important, magnetic metal Ni appropriate for accurate description of finite temperature and defect (point defect, surface energies, stacking fault, dislocation core, grain boundary) properties and phase transformations.
The new potential provides a robust, predictive tool for the study of Ni and its mechanical behavior.

\section{Results and Discussion}

\label{sec:Results and Discussion}

The DP model for Ni (DP-Ni) is trained via a supervised ML technique. 
The training labels include atom coordinates, total energy, atomic forces, and virial tensors, obtained from spin-polarized DFT calculations. 
We employ the DP-GEN framework~\cite{zhang2020dp} along with the Deep Potential Smooth Edition (DeepPot-SE)~\cite{zhang2018end} to conduct the training. 
A ``specialization'' strategy~\cite{wen2021specialising} is adopted to further improve the accuracy. 
Initially, distorted $2 \times 2 \times 2$ body-centered cubic (BCC), FCC, and HCP structures are input into finite-temperature $ab$ $initio$ molecular dynamics (AIMD) simulations to generate a starting training dataset (108 entries). 
During the DP-GEN loop, exploration involves DP-based MD (DPMD) simulations on bulk and surface structures for several temperatures and pressures, followed by DFT calculations on selected configurations. 
The resultant DFT data is then incorporated into the training dataset to refine the DP models. 
Convergence of the DP-GEN loop is achieved when the agreement between DP and  DFT calculations for atomic forces reaches a predetermined threshold.

Following the DP-GEN loop, the resulting DP model can accurately represent the general properties of FCC and HCP Ni albeit with some discrepancies in the cohesive energy curve compared to DFT results. 
To address this, specialized training datasets are generated from selected configurations along the cohesive energy line. 
These specialized training datasets are then merged with those generated from the DP-GEN loop. 
The combined training dataset used for potential development consists of 2,020 entries, all derived from spin-polarized DFT calculations. For a comprehensive discussion on the training process and training data generation, please refer to Section~\ref{sec:Training Strategy}.

We systematically benchmark a wide range of crystal and defect properties of DP-Ni; in particular, we examine equations of states, elastic constants, finite temperature properties, phonon spectra, point defect energies, surface properties, stacking fault energies, plastic deformation, dislocation dissociation, and grain boundary energies. 
We compare the DP-Ni model performance against several of the most widely-used and best-performed empirical/semi-empirical interatomic potentials, including the EAM potential of Mishin $et~al$.~\cite{mishin1999interatomic}, 
the MEAM\_2021 potential of Vita $et~al$.~\cite{vita2021exploring},
the MEAM\_2015 potential of  Ko $et~al$.~\cite{ko2015development}, and the ML qSNAP potential by Zuo $et~al$~\cite{zuo2020performance}. 
These benchmarks provide a comprehensive assessment of the performance of our new DP-Ni model with other widely-used interatomic potentials for Ni.

\subsection{Basic Crystal Properties}

Table~\ref{tab:basic properties} compares a wide range of crystalline Ni properties with DFT calculations, experiments, DP-Ni, and other interatomic potentials. 
The DP-Ni shows excellent agreement with both DFT and experimental values for the stable FCC and metastable HCP crystals. 
The energy difference between DP-Ni and DFT is within 3 meV/atom for both FCC and HCP Ni, while the lattice parameter difference between DP and DFT/experiment is within 0.004 \AA. 
The EAM and MEAM potentials also exhibit accurate lattice parameters for both phases, with discrepancies less than 2\% when compared to DFT and experimental results. 
The DP-Ni model shows a slight deviation of the cohesive energy from the experimental data but accurately reproduces the DFT value for FCC Ni (this is likely associated with issues related to the DFT data to which DP-Ni is trained). 
EAM and MEAM\_2015 potentials perfectly match the experimental cohesive energy of 4.450 eV/atom as required in their fitting procedure, while MEAM\_2021 underestimates it by $\sim$11\%. 
In contrast, qSNAP potential exhibits a large deviation $\sim$30\% from the experimental data. 
DP-Ni yields cohesive energy that is almost identical to the DFT prediction for HCP Ni. 
Similarly, EAM and MEAM\_2015 yield results close to the experimental measurements, while the other interatomic potentials exhibit significant deviations from both DFT and experimental results.

\begin{table*}[!htbp]
  \caption{\label{tab:basic properties}
  Comparison of several crystal properties obtained from DFT, experiment (Expt.), and various interatomic potentials (DP-Ni, EAM~\cite{mishin1999interatomic}, MEAM\_2021~\cite{vita2021exploring}, MEAM\_2015~\cite{ko2015development},  qSNAP~\cite{zuo2020performance}); i.e.,  
  lattice parameters ($a$), bulk energies ($E$), cohesive energies ($E_\mathrm{coh}$), elastic constants ($C_{ij}$) of FCC and HCP Ni, and FCC melting point. 
  Bold numbers indicate deviations of $>15$\% versus DFT and/or experiment.
  }
\begin{ruledtabular}
\begin{tabular}{ccccccccc}
\textrm{Structure}&
\textrm{Property}&
\textrm{DFT}&
\textrm{Expt.}&
\textrm{DP}&
\textrm{EAM}&
\textrm{MEAM\_2021}&
\textrm{MEAM\_2015}&
\textrm{qSNAP}\\
\colrule
        
  FCC & $a$ (Å) &3.517 &3.520\footnote{Lattice constants at 6 K~\cite{kanhe2016investigation}. $^\text{b}$ \cite{kittel2005introduction}. $^\text{c}$ Experimental elastic constants at 0 K  extrapolated from low T data~\cite{simmons1971single}. $^\text{d}$ \cite{dinsdale1991sgte}. $^\text{e}$ Lattice constants $a$ and $c/a$ ratio at room temperature~\cite{lagrow2013can}. $^\text{f}$ $E_\mathrm{coh}$ of FCC based on the DFT energy difference between FCC and HCP.}  &3.518 &3.520 &3.519 &3.521 &3.521\\
& $E$ (eV/atom)&-5.467 &- &-5.466 &-4.450 & -3.952 & -4.450 &-5.780\\
& $E_\mathrm{coh}$ (eV/atom)&4.865 &4.450$^\text{b}$ &4.862 &4.450 &3.952 &4.450 &\textbf{5.780}\\
& $C_{11}$ (GPa)&275.7 &261.2$^\text{c}$ &278.9 &247.9 &278.3 &260.4 &267.5\\
& $C_{12}$ (GPa)&156.0 &150.8$^\text{c}$ &158.1 &147.8 &169.8 &148.6 &155.3\\
& $C_{44}$ (GPa)&130.7 &131.7$^\text{c}$ &127.7 &124.8 &112.5 &\textbf{111.1} &125.7\\
& $T_\mathrm{m}$ (K)&-&1728$^\text{d}$&1635 &- &- &1892 &-\\
&&&&&&&\\  
HCP & $a$ (Å)&2.484&2.487$^\text{e}$ & 2.485 & 2.483 &2.490 &2.487 &2.491\\
& $c/a$ &1.643 &1.645$^\text{e}$ &1.641 &1.619 &1.630 &1.642 & 1.643\\
& $E$ (eV/atom)&-5.443 &- &-5.446 &-4.430 &-3.956 &-4.440 &-5.772\\
& $E_\mathrm{coh}$ (eV/atom)&4.841 &4.426$^\text{f}$ &4.842 & 4.430 & 3.956 & 4.440 & \textbf{5.772}\\
& $C_{11}$ (GPa)&312.0 &- &311.4 &302.2 &327.6 &314.7 &334.0\\
& $C_{12}$ (GPa)&142.3 &- &156.0 &147.6 &159.5 &133.8 &144.0\\
& $C_{13}$ (GPa)&122.8 &- &114.6 &\textbf{76.9}  &131.9 &108.3 &109.1\\
& $C_{33}$ (GPa)&330.7 &- &344.7 &\textbf{213.3} &355.6 &336.0 &369.2\\
& $C_{44}$ (GPa)&55.5  &- &54.6  &\textbf{64.3}  &\textbf{73.0}  &\textbf{77.2}  &\textbf{77.2}\\
        
\end{tabular}
\end{ruledtabular}
\end{table*}

Elastic constants are fundamental and essential material properties reflecting mechanical stability and stiffness. 
The largest discrepancy between DP and DFT/experiment for DP FCC Ni is for $C_{44}$ (2.3\%)/$C_{11}$ (6.8\%); all other interatomic potentials also accurately reproduce the elastic constants of FCC Ni (except for a slight underestimation of $C_{44}$ for the MEAM potentials). 
For HCP Ni, the predicted $C_{ij}$s from DFT and DP yield mechanical stability according to the Born criteria~\cite{Mouhat_Prb}; i.e.,  $C_{11} -|C_{12}|>0$, $(C_{11}+C_{12})C_{33}-2C_{13}^2>0$ and $C_{44}>0$. 
DP-Ni accurately reproduces the DFT elastic constants of HCP Ni, with a maximum deviation at $C_{12}$ (9.6\%). 
On the other hand, all other potentials show large deviations in the elastic constants of HCP Ni as compared with DFT results; particularly for EAM at $C_{13}$ (37.4\%), $C_{33}$ (35.5\%), and $C_{44}$ (15.9\%), and the other three potentials at $C_{44}$ 31.5\%, 39.1\%, 39.1\% for MEAM\_2021, MEAM\_2015, and qSNAP, respectively. 
The elastic constants measure the (stress) response of the crystal to small strains and are indicative of sensitivity to lattice distortions. 
The training data for the DP-Ni includes many such locally distorted structures (see Section~\ref{sec:Training Strategy}). 
No HCP crystal distortions are included in the fitting data of the classical and ML qSNAP potentials.

\subsection{Phonon Spectra}

\begin{figure*}[!htbp]
\includegraphics[width=1\textwidth]{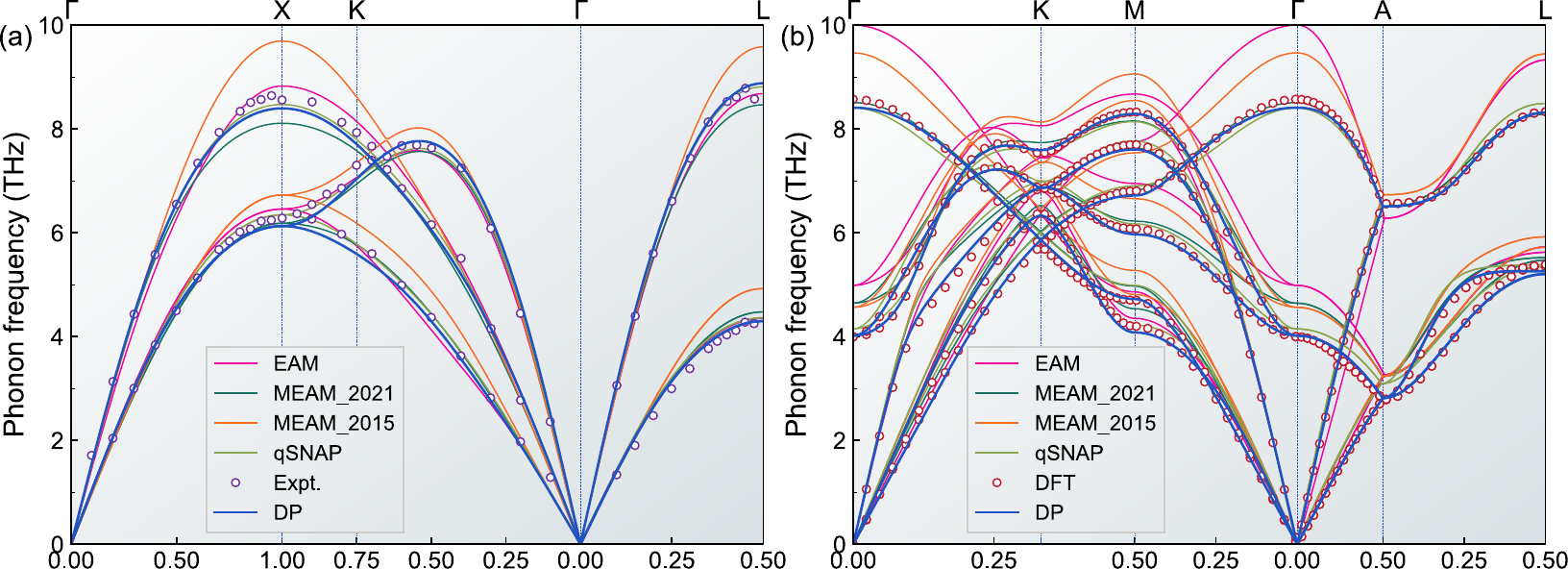}
\caption{\label{fig:phonon}Comparison of predicted and measured phonon spectra for (a) FCC and (b) HCP. The experimental values are from FCC Ni neutron diffraction data at 298 K~\cite{birgeneau1964normal}, and the HCP DFT data is from this work.}
\end{figure*}

In addition to the Born mechanical stability criteria and cohesive energies, phonon spectra~\cite{grimvall2012lattice} also characterize crystal stability. 
Figure~\ref{fig:phonon} shows the phonon spectra of both FCC and HCP Ni obtained from experiment~\cite{birgeneau1964normal}, DFT, DP-Ni, and other interatomic potentials. Both FCC and HCP Ni are inherently stable (no imaginary frequencies). 
However, notable variations in accuracy are observed amongst the different potentials. 
DP-Ni demonstrates outstanding performance in both FCC and HCP crystal structures, reproducing all frequencies across the phonon spectra with high accuracy. 
Minor deviations are observed for qSNAP potential (particularly the HCP). 
Other classical potentials exhibit evident deviations from the DFT and/or experimental data at symmetry points. 
                
\subsection{FCC Surface Energies and Point Defects}

\begin{table*}[!htbp]
\caption{\label{tab:surface and defects energies}
The calculated unrelaxed surface energies ($E_{\mathrm{s}}$), vacancy formation energies ($E_{\mathrm{v}}^{\mathrm{f}}$), interstitial formation energies ($E_{\mathrm{i}}^{\mathrm{f}}$), and unstable ($\gamma_{\mathrm{usf}}$) and stable stacking fault energies ($\gamma_{\mathrm{sf}}$), as well as grain boundary energies of low Miller index tilt boundaries for FCC Ni using DP-Ni, in comparison with DFT results, available experimental data, and selected interatomic potentials.  
$\circledast$ indicates that the initial interstitial structure is not stable and will undergo a transformation to the $\langle100\rangle$ dumbbell. 
Bold numbers indicate $>15$\% deviations from DFT/Expt.}
\begin{ruledtabular}
\begin{tabular}{cccccccc}
\textrm{Property}&
\textrm{DFT}&
\textrm{Expt.}&
\textrm{DP}&
\textrm{EAM}&
\textrm{MEAM\_2021}&
\textrm{MEAM\_2015}&
\textrm{qSNAP}\\
\colrule            
$E_{\mathrm{s}}${\{111\}} (J/m\textsuperscript{2}) &1.919 &\multirow{6}{*}{2.240\footnote{Polycrystalline average~\cite{tyson1977surface}. $^\text{b}$ \cite{megchiche2006density}. $^\text{c}$ Variant $\langle100\rangle$ dumbbell. $^\text{d}$ \cite{carter1977stacking,murr1975}.
}} &1.958 &1.636 &1.815 &\textbf{1.630} &1.938\\
$E_{\mathrm{s}}${\{221\}} (J/m\textsuperscript{2}) &2.210 &     &2.259 &1.924 &2.164 &1.965 &2.230\\
$E_{\mathrm{s}}${\{110\}} (J/m\textsuperscript{2}) &2.343 &     &2.357 &2.056 &2.367 &2.172 &2.356\\
$E_{\mathrm{s}}${\{211\}} (J/m\textsuperscript{2}) &2.279 &     &2.323 &1.970 &2.222 &2.021 &2.280\\
$E_{\mathrm{s}}${\{210\}} (J/m\textsuperscript{2}) &2.463 &     &2.488 &2.181 &2.526 &2.321 &2.472\\
$E_{\mathrm{s}}${\{100\}} (J/m\textsuperscript{2}) &2.239 &     &2.223 &\textbf{1.884} &2.220 &2.088 &2.254\\
\\
$E_{\mathrm{v}}^{\mathrm{f}}$ (eV) &1.424 &1.400--1.800$^\text{b}$     &1.236 &1.598 &1.539 &1.509 &1.465\\

$E_{\mathrm{i}}^{\mathrm{f}}$ $\langle100\rangle$ dumbbell (eV) &4.048 &- &4.184 &\textbf{4.885}$^\text{c}$ &4.253 &4.531 &4.118\\
$E_{\mathrm{i}}^{\mathrm{f}}$ $\langle111\rangle$ dumbbell (eV) &4.664 &- &4.892 &\textbf{6.920} &4.765 &\textbf{5.508} &4.751\\
$E_{\mathrm{i}}^{\mathrm{f}}$ $\langle110\rangle$ dumbbell (eV) &4.828 &-     &4.614 &\textbf{5.786} &4.664 &5.103 &4.769\\
$E_{\mathrm{i}}^{\mathrm{f}}$ Crowdion (eV)                     &4.826 &-     &4.614 &5.114 &4.669 &5.112 &4.788\\
$E_{\mathrm{i}}^{\mathrm{f}}$ Octahedral (eV)                   &4.229 &-     &4.421 &$\circledast$ &4.465 &$\circledast$  &4.460\\
$E_{\mathrm{i}}^{\mathrm{f}}$ Tetrahedral (eV)                  &4.670 &-     &4.986 &\textbf{6.920} &5.085 &\textbf{5.508} &$\circledast$\\
\\
$\gamma_{\mathrm{usf}}$ $\langle110\rangle$ (mJ/m\textsuperscript{2}) &766.6 &- &801.6 &\textbf{924.3} &746.9 &\textbf{898.2} &789.9\\
$\gamma_{\mathrm{usf}}$ $\langle112\rangle$ (mJ/m\textsuperscript{2}) &280.4 &- &301.9 &\textbf{365.6} &285.4 &\textbf{423.6} &275.5\\
$\gamma_{\mathrm{sf}}$ $\langle112\rangle$ (mJ/m\textsuperscript{2}) &135.9 &125$^\text{d}$ &126.8 &125.2 &\textbf{-26.9} &\textbf{60.0} &\textbf{52.2}\\
\\
$\Sigma$3 $[1\bar{1}0]$ $(111)$ (mJ/m\textsuperscript{2})  &68.03  &- &63.50 &63.46 &\textbf{-13.45}  &\textbf{30.09} &\textbf{26.53}\\
$\Sigma$3 $[1\bar{1}0]$ $(112)$ (mJ/m\textsuperscript{2}) &896.03 &- &893.67 &\textbf{1064.03} &782.53 &960.66 &908.44\\
$\Sigma$5 $[100]$ $(0\bar{2}1)$ (mJ/m\textsuperscript{2}) &1288.75 &- &1310.72 &\textbf{1564.08} &1372.11 &1421.66 &1339.00\\
$\Sigma$7 $[111]$ $(3\bar{2} \bar{1})$ (mJ/m\textsuperscript{2})  &1234.31 &- &1212.57 &\textbf{1471.91} &1210.14 &1395.51 &1286.36\\
$\Sigma$9 $[1\bar{1}0]$ $(22\bar{1})$ (mJ/m\textsuperscript{2}) &1120.58 &- &1103.69 &\textbf{1368.13} &1148.89 &1258.83 &1157.30\\
$\Sigma$11 $[1\bar{1}0]$ $(113)$ (mJ/m\textsuperscript{2})  &454.23 &- &440.81 &\textbf{531.15} &420.36  &518.89 &464.21\\
                
\end{tabular}
\end{ruledtabular}
\end{table*}

In Table~\ref{tab:surface and defects energies}, the unrelaxed energies of low Miller index surfaces calculated by DP-Ni are compared to values obtained from DFT, experiments, and other potentials. 
Our DFT results are consistent with both the values and ordering of previous DFT calculations~\cite{tran2016surface}. 
DFT predicts that the \{111\} close-packed plane has the lowest surface energy, whilst the \{210\} surface is the highest. 
All interatomic potentials successfully reproduce the lowest and highest energy planes. 
DP-Ni results show excellent agreement with DFT, exhibiting a maximum error of 2.2\% for the \{221\} surface. 
MEAM\_2021 and qSNAP potentials also provide accurate predictions (errors within 6\%). 
However, EAM and MEAM\_2015 potentials slightly underestimate surface energies by 11.4\%-15.9\% and 5.8\%-15.1\%, respectively. 
The quoted experimental surface 2.240 J/m\textsuperscript{2} is a polycrystalline average ~\cite{tyson1977surface}. 
The vacancy formation energy ($E_{\mathrm{v}}^{\mathrm{f}}$) from DP-Ni is $\sim$0.124 eV which is 13.2\% lower than the DFT value. 
All other potentials yield higher values than the DFT $E_{\mathrm{v}}^{\mathrm{f}}$ result.

\begin{figure}[!htbp]
  \includegraphics[width=0.48\textwidth]{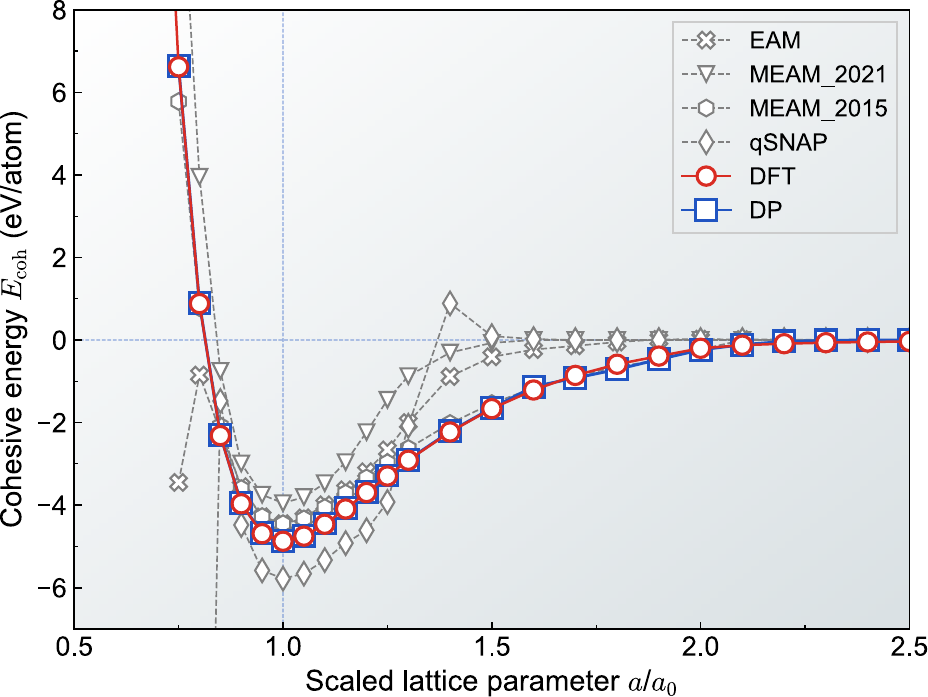}
  \caption{\label{fig:cohesive_pt}The FCC Ni cohesive energy as a function of lattice parameter from DFT and several potentials. 
  }
\end{figure}

\begin{figure*}[!thbp]
  \includegraphics[width=0.9\textwidth]{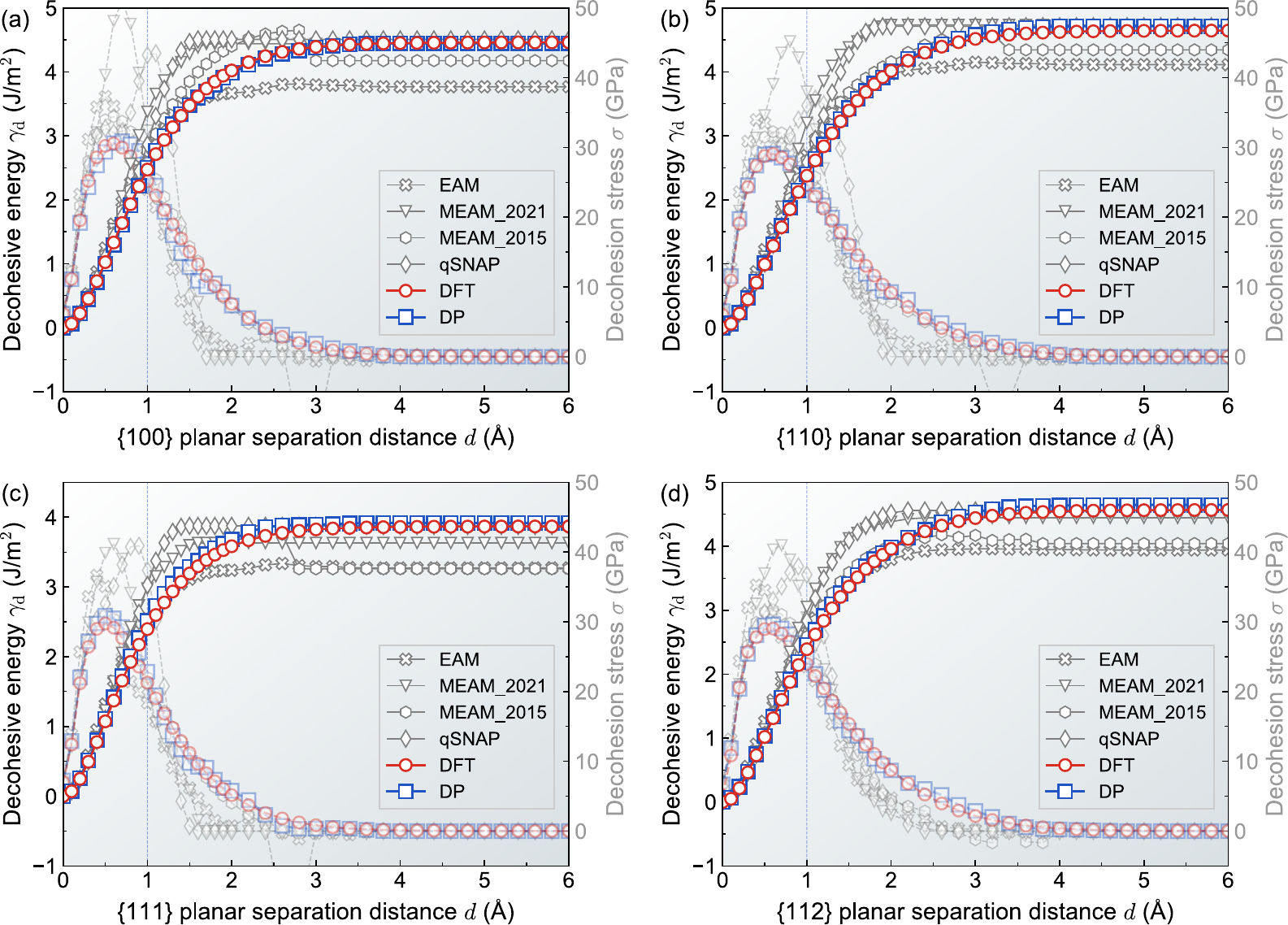}
  \caption{\label{fig:decohesive}The plane decohesion energy ($\gamma_\mathrm{d}$ - bright data points) and stress ($\sigma$ - dim data points) as a function of plane separation distance ($d$) for (a) $\{100\}$, (b) $\{110\}$, (c) $\{111\}$ and (d) $\{112\}$ planes.}
  \end{figure*}

The FCC structure exhibits six types of self-interstitial structures, namely the $\langle100\rangle$ dumbbell, $\langle111\rangle$ dumbbell, $\langle110\rangle$ dumbbell, crowdion, octahedral, and tetrahedral (see Supplementary Fig.~S1). 
The DFT calculations show that $\langle100\rangle$ dumbbell has the lowest formation energy in FCC, followed by octahedral, $\langle111\rangle$ dumbbell, tetrahedral, crowdion and $\langle110\rangle$ dumbbell. 
Note that the crowdion and $\langle110\rangle$ dumbbell energies are nearly equivalent, and relaxed configurations exhibit a slight difference. 
This energy ordering is consistent with other results~\cite{toijer2021solute,tucker2009determination,ma2021nonuniversal}. 
DP-Ni captures all of the metastable configurations with a maximum energy discrepancy of $<6.8$\% (tetrahedral) compared to DFT results. 
However, a small inconsistency with DFT is the altered energy ordering sequence for DP, which is (from low to high): $\langle100\rangle$ dumbbell, octahedral, crowdion, $\langle110\rangle$ dumbbell, $\langle111\rangle$ dumbbell and tetrahedral. 
Almost all EAM potential self-interstitial energies are much higher than the DFT values, the octahedral interstitial is unstable, and the EAM $\langle100\rangle$ dumbbell is short. 
The MEAM\_2021 captures all six self-interstitial configurations with small energy deviation compared to DFT, but the energy ordering is quite different. 
The $\langle111\rangle$ dumbbell and tetrahedral energies from MEAM\_2015 are nearly the same after relaxation;  the octahedral structure transforms into a $\langle100\rangle$ dumbbell. 
The qSNAP potential accurately reproduces all self-interstitial formation energies; however, the tetrahedral interstitial transforms to a $\langle100\rangle$ dumbbell.

Note that the training datasets for the DP-Ni potential do not include vacancy or self-interstitial configurations. 
This implies that DP-Ni accurately captures the essential characteristics of many defects in Ni even though such configurations are not included in the training data. 
This underscores the versatility and reliability of the DP-Ni model in predicting defect properties. 

\subsection{Cohesive and Decohesive Energy}

\begin{figure*}[!htbp]
\includegraphics[width=0.9\textwidth]{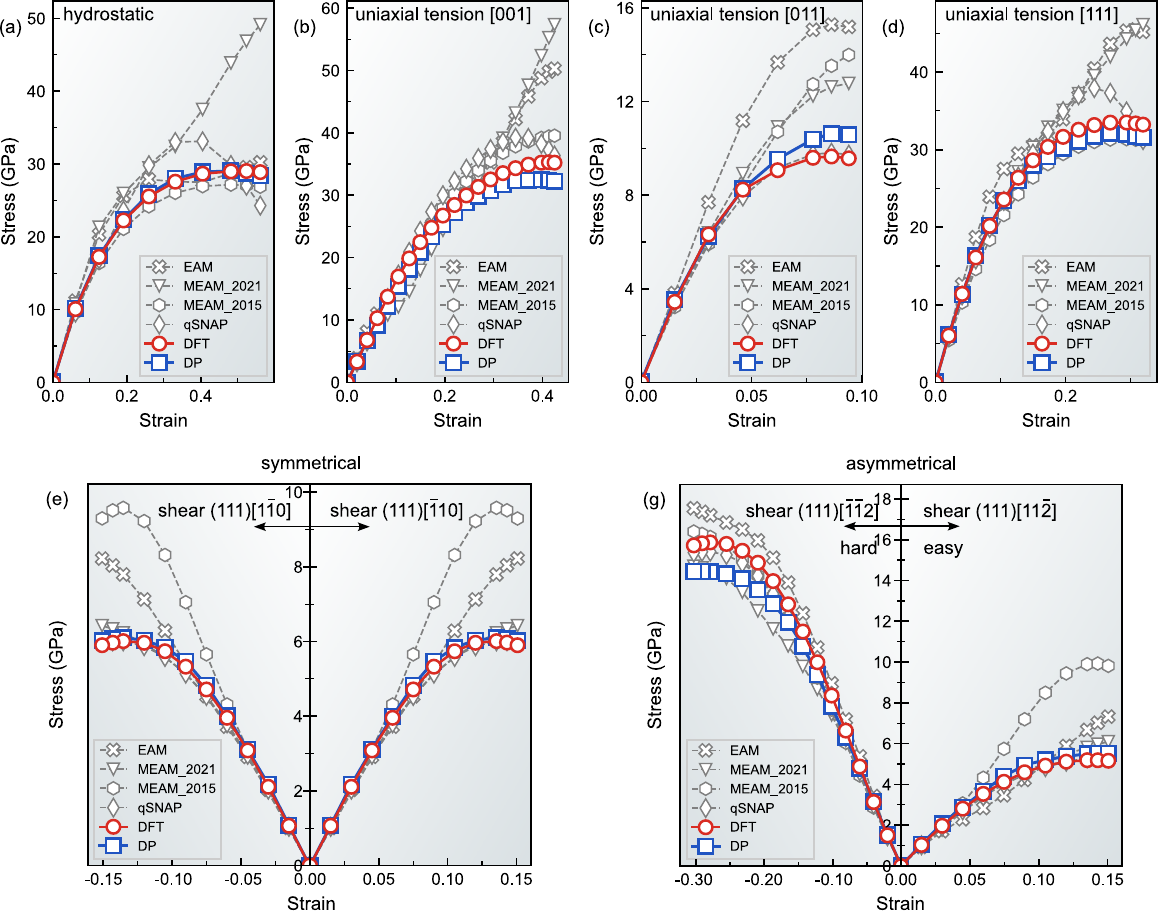}
\caption{\label{fig:ids} Stress-strain curves for (a) hydrostatic, (b) $[001]$, (c) $[011]$, (d) $[111]$ uniaxial tension and (e) $(111)[1\bar{1}0]$, $(111)[\bar{1}10]$, (f) $(111)[\bar{1}\bar{1}2]$, $(111)[11\bar{2}]$ shear loading.}
\end{figure*} 

The relationship between the cohesive energy and atomic spacing (cohesion curves) is critical for a wide range of properties. 
Figure~\ref{fig:cohesive_pt} shows the cohesive curves for FCC Ni at 0 K, determined from DFT and interatomic potentials. 
The DFT, DP-Ni, MEAM\_2021, and MEAM\_2015 curves are smooth across the entire range. 
The DP-Ni and DFT curves nearly overlap, while the MEAM\_2015 deviates from the DFT value near the equilibrium lattice parameter. 
In contrast, the MEAM\_2021 results show large deviations from the DFT data in the crucial 0.5$a_0$ to 2$a_0$ range. 
The EAM curve remains continuous at large atom separations but exhibits discontinuities under large compression, with deviations from the DFT curve in the $1.25-2.0 a_0$ range. 
The qSNAP potential yields discontinuous and inaccurate cohesive energy curves and its equilibrium FCC Ni cohesive energy is substantially different from the DFT results (see Table~\ref{tab:basic properties}). 
This indicates that the qSNAP potential may introduce unexpected and significant errors in mechanical properties.

Examining the (uniaxial) surface decohesion energy and its gradient (stress) provides a deeper understanding of the energy landscape and forces involved in atomic plane separation; this is important for predicting and simulating fracture. Figure~\ref{fig:decohesive} displays the surface decohesion energy and its gradient for four crystallographic planes using DFT and interatomic potentials. 
No plane separation data is explicitly included in the DFT training datasets of DP-Ni. 
The DP-Ni model demonstrates excellent predictability compared with DFT for all planes. 
The MEAM\_2021 and qSNAP potentials also show relatively good agreement with DFT data. 
However, significant deviations in energy and stress are observed for separation distance ranging from $0.5<d<3$~{\AA} for the \{100\}, \{110\}, \{112\} planes and $0.5<d<2.5$~{\AA} for the \{111\} plane. 
Additionally, the peak stress position for qSNAP is shifted to larger $d$ ($\sim1$~{\AA}). 
The EAM potential exhibits discontinuities for \{100\}, \{110\}, \{111\} planes for $1.5<d<3.5$~{\AA}, leading to  unphysical fluctuating decohesion stresses. 
The EAM decohesion energies are much smaller than DFT for $d>1.5$~{\AA}. 
Peak stress values for the EAM potential are shifted to smaller $d$.  
The MEAM\_2015 potential yields decohesion results largely in agreement with DFT results except for abrupt jumps at $d\sim2.5$~{\AA}.

\subsection{Ideal Strength}

Smooth cohesive and decohesive energies are important for predicting (ideal) strength. 
Ideal strength is the maximum stress that a perfect material can withstand before undergoing plastic deformation or fracture~\cite{jhi2001mechanical}. 
This property can be identified through the stress-strain curve (a valuable tool for material application and design). 
We initially assess the ideal strength of FCC Ni under tensile and shear loading using DFT calculations; see Fig.~\ref{fig:ids} for the computed stress as a function of applied strain in various directions. 
At low strains, the curves are linear (linear elastic), while at higher strains the deviation from the linear elastic response is evident; the ideal strength ($\sigma_\mathrm{ideal}$) corresponds to the maximum stress or the stress at the peak strain ($\epsilon_\mathrm{ideal}$). 
The stress-strain response is strongly anisotropic. 
For example, the $\sigma_\mathrm{ideal}$ and $\epsilon_\mathrm{ideal}$ differ considerably between the $[001]$ and $[011]$ directions under  uniaxial tension. 
Additionally, the $(111)\langle 112 \rangle$ directions under shear stress show obvious ``stiff'' and ``soft'' tendencies. 
Overall, Ni shows $\sigma_\mathrm{ideal} = 29.0$ GPa and $\epsilon_\mathrm{ideal}=0.52$ under  hydrostatic tension while $\sigma_\mathrm{ideal}= 35.3$ GPa and $\epsilon_\mathrm{ideal}=0.41$ in $[001]$ uniaxial tension and 
$\sigma_\mathrm{ideal} = 15.9$ GPa and $\epsilon_\mathrm{ideal}=0.28$ in  $(111)[\bar{1}\bar{1}2]$ shear.

\begin{figure*}[!htbp]
  \includegraphics[width=0.96\textwidth]{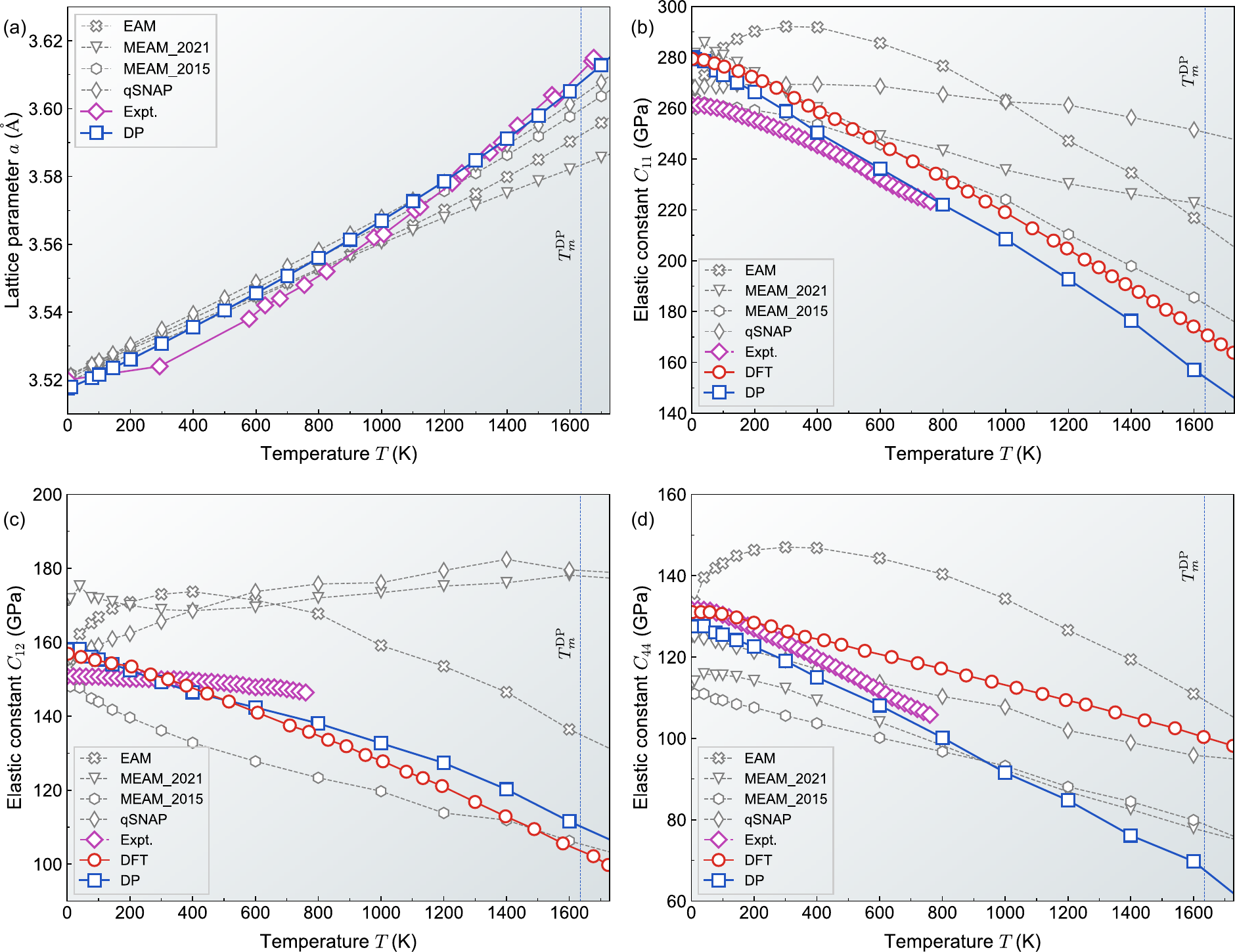}
  \caption{\label{fig:cij_T}Finite-temperature properties of FCC Ni calculated from DFT, DP-Ni and other potentials. (a) lattice parameters and elastic constants (b) $C_{11}$, (c) $C_{12}$ and (d) $C_{44}$. The experimental values of lattice parameters are from \cite{suh1988high}, while the DFT and experimental data of elastic constants are from \cite{HACHET2018280} and~\cite{ALERS196040}, respectively. 
  The DP melting point $T^\text{DP}_m$ is indicated by vertical dotted lines ($\sim90$ K lower than the experiment - termination of the temperature axis on the plots)}
  \end{figure*}

Unlike elastic constants, ideal strength calculations involve significant (rather than infinitesimal) deformation and therefore represent considerable demands on the ability of a potential to accurately describe deformation. 
We conduct a comparative analysis of stress-strain relationships among various interatomic potentials using static calculations.
The DP-Ni is in excellent agreement with the DFT results, especially for hydrostatic and $[111]$ uniaxial tension, as well as $(111)[1\bar{1}0]$, $(111)[\bar{1}10]$ and $(111)[11\bar{2}]$ shear. 
The largest deviations observed are 10.4\% and 9.3\% in the non-linear region for $[011]$ tension and $(111)[\bar{1}\bar{1}2]$ shear, respectively. 
The DP-Ni also reproduces the $\epsilon_\mathrm{ideal}$ in all cases. 
MEAM\_2021 performs well in shear but overestimates the ideal strength and strain in hydrostatic and uniaxial tension. 
Similarly, qSNAP shows good performance in shear but overestimates $\sigma_\mathrm{ideal}$ and underestimates $\epsilon_\mathrm{ideal}$ under hydrostatic and most uniaxial tension cases. 
The EAM model shows large deviations compared to DFT results, with a discrepancy of 58.7\% under $[011]$ tension. 
In comparison with DFT,  MEAM\_2015 exhibits minor discrepancies in hydrostatic but overestimates the ideal strength in most uniaxial tension and shear cases.

\subsection{Finite Temperature Properties}

\begin{figure*}[!htbp]
\includegraphics[width=0.95\textwidth]{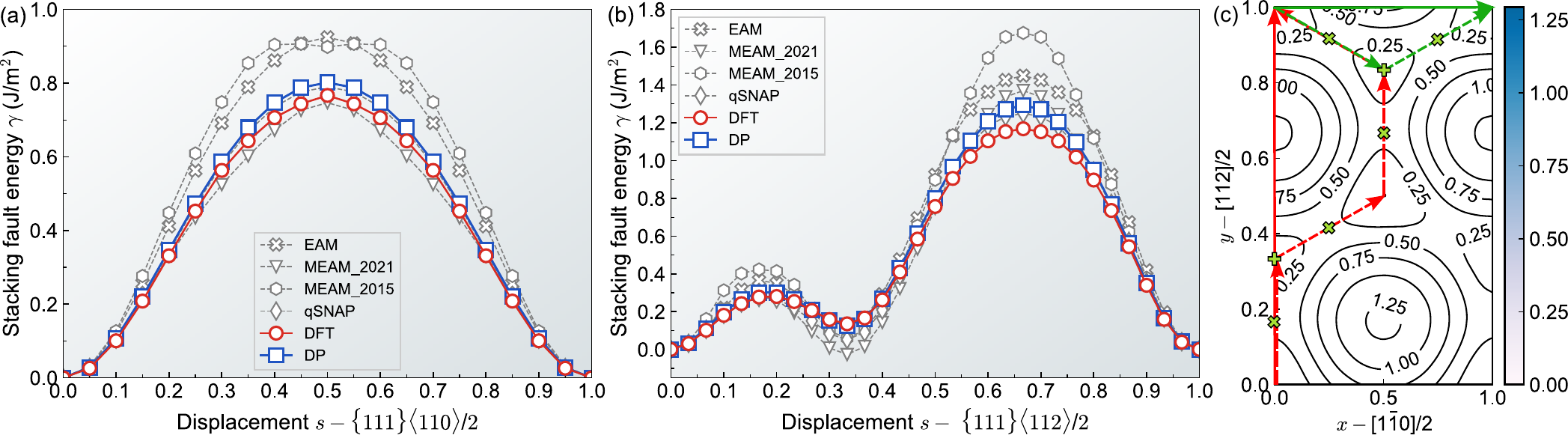}
\caption{\label{fig:gamma_line_surface} Generalized stacking fault energy (GSFE) lines ($\gamma$-lines) along the (a) \adisl and (b) \bdisl directions, as well as (c) GSFE surface ($\gamma$-surface) on the loosest packing $\{111\}$ planes predicted by DP-Ni. The solid red and green arrows represent the slip path along \bdisl and \adisl directions respectively, and dashed arrows show the corresponding dissociated slip paths. The symbols of $+$ and $\times$ represent the positions of stable and unstable stacking faults.}
\end{figure*}

Nickel and Nickel-based alloys are widely used at elevated temperatures, such as in superalloy turbine blades, hence we also focus on the finite temperature properties using DP-Ni in MD simulations. 
Figure~\ref{fig:cij_T} shows the variation of the FCC Ni lattice parameter and elastic constants as compared with experimental measurements~\cite{suh1988high} and simulations with other interatomic potentials from 0 to 1700 K. 
The DP-Ni lattice parameter is in good agreement with experimental results at high temperatures (above 600 K)~\cite{suh1988high} and the thermal expansion coefficient (slope) is similar to the experimental value. 
The DP-Ni melting point for Ni is $1635\pm5$ K (Table~\ref{tab:basic properties}), obtained using the two-phase method~\cite{Morris_Prb},
is $\sim$5.4\% lower than the experimental value (1728 K). 
Figures~\ref{fig:cij_T}(b)-(d) show the temperature dependence of the elastic constants $C_{ij}$ from DFT  within the quasi-harmonic approximation~\cite{HACHET2018280} 
and various potentials. 
Like the DFT and experimental results~\cite{HACHET2018280,ALERS196040}, the DP-Ni elastic constants decrease continuously with temperature. 
Other potentials reveal different trends or profiles as compared with DFT/experiment. 
The EAM data shows an increase of $C_{ij}$ with temperature below 400 K, followed by a continuous decrease at higher temperatures. 
This abnormal elastic constant behavior is also observed for MEAM\_2021 and qSNAP for $C_{12}$. 
On the other hand, the MEAM\_2015 elastic constants results show a similar (decreasing) trend as the DFT/experiment results, although the discrepancies in the magnitude can be significant, e.g., the discrepancy is $>15$\% for $C_{12}$ for $T>700$ K. 
The present results demonstrate that most potentials are unreliable for predicting finite temperature behavior.  This is likely because they were fitted to low temperature (and/or limited finite temperature) data, while the DP method incorporates finite-temperature-like perturbations in the training set. The discrepancies between the finite-temperature DFT  and experimental results may be attributed to several factors. 
These include issues related to the exchange-correlation function~\cite{HACHET2018280} and approximations employed in extracting finite temperature results from DFT calculations (e.g., the quasi-harmonic approximation).

\subsection{Stacking Fault and Dislocation Core}

The generalized stacking fault energy (GSFE) is a useful, surrogate property for predicting the plastic response of the material, i.e., dislocation and twinning properties~\cite{xiao_2023_mt}. 
The GSFE represents the variation in the system energy required for the slip of a part of the crystal over the other along particular crystal lattice planes under shear, leading to the formation of stacking faults. 
The variation of the system energy accompanying the translation/slip along particular directions on a slip plane is referred to as the $\gamma$-line~\cite{christian1970dislocations}. 
The maximum energy along the $\gamma$-line corresponds to the unstable stacking fault energy ($\gamma_\mathrm{usf}$), which represents the barrier for dislocation nucleation at stress concentrations such as crack tips. 
A metastable point on the $\gamma$-line, referred to as $\gamma_\mathrm{sf}$, represents a dislocation dissociation energy. 
The complete two-dimensional plane characterizing all possible slip directions, $\gamma$-lines, is the $\gamma$-plane or $\gamma$-surface~\cite{christian1970dislocations}.

Figures~\ref{fig:gamma_line_surface}(a)-(b) show the $\gamma$-lines along the \adisl and \bdisl directions on the $\{111\}$ plane (most dense plane in FCC) determined from DFT and several interatomic potentials. 
The only unstable stacking fault is along the \adisl-direction with a $\gamma_\mathrm{usf}$ of 766.6 mJ/m\textsuperscript{2} at half of the Burger vector \textbf{b}, \andisl, consistent with previous DFT calculations~\cite{su2019density}. 
In the \bdisl slip direction, a stable stacking fault occurs at \textbf{b}/3 (\textbf{b}=\bndisl) while an unstable stacking fault is present at \textbf{b}/6. 
Although a peak appears at 2\textbf{b}/3 in the $\gamma$-line, it is irrelevant because this barrier, 1168.4 mJ/m\textsuperscript{2}, is too high to allow slip. 
The DFT calculations yield $\gamma_\mathrm{usf}=280.4$ mJ/m\textsuperscript{2} and $\gamma_\mathrm{sf}=135.9$ mJ/m\textsuperscript{2}. 
The $\gamma_\mathrm{sf}$ is in good agreement with experimental results (125 mJ/m\textsuperscript{2}~\cite{carter1977stacking,murr1975}) and previous DFT values ranging from 110 to 145 mJ/m\textsuperscript{2}~\cite{su2019density,rodney2017ab}. Figures~\ref{fig:gamma_line_surface} (a) and (b) also show the $\gamma$-line results from DP-Ni and other interatomic potentials. 
All potentials reproduce the general shape of the $\gamma$-lines from DFT except for MEAM\_2015, which shows a minimum value at \textbf{b}/2 along \adisl direction. Table~\ref{tab:surface and defects energies} lists the calculated $\gamma_\mathrm{usf}$ and $\gamma_\mathrm{sf}$ values. 
DP-Ni reproduces the different stacking fault energies well compared with DFT results, with deviations of only 4.6\% for $\gamma_\mathrm{usf}$ in the \adisl slip, 7.6\% for $\gamma_\mathrm{usf}$ and 6.7\% for $\gamma_\mathrm{sf}$ along the \bdisl slip. (Notably, there is no stacking fault data in the DP-Ni training datasets.) 
In contrast, the MEAM\_2021 and qSNAP potentials capture the $\gamma_\mathrm{usf}$ well in both \adisl and \bdisl directions, but significantly underestimate the $\gamma_\mathrm{sf}$, particularly for MEAM\_2021 which yields an unphysical negative $\gamma_\mathrm{sf}$. 
While the EAM potential accurately describes the $\gamma_\mathrm{sf}$, it overestimates both $\gamma_\mathrm{usf}$ in \adisl and \bdisl directions. 
The MEAM\_2015 potential fails to accurately describe $\gamma_\mathrm{usf}$ and $\gamma_\mathrm{sf}$. 
The unrealistic empirical and ML qSNAP potential GSFE results suggest that these potentials will struggle to correctly simulate dislocation nucleation and dislocation dissociation behavior. 
The minimum energy path is indicated on the DP-Ni $\{111\}$ $\gamma$-surface (Fig.~\ref{fig:gamma_line_surface}(c)), which exhibits the expected symmetry from geometry.  
The minimum energy path for dislocation dissociation is expected to follow the green or red dashed arrows (Fig.~\ref{fig:gamma_line_surface}(c)), indicating that a full dislocation \adisl or \bdisl will dissociate into Schockley partials on the $\{111\}$ plane - as expected.

\begin{figure*}[!htbp]
\includegraphics[width=0.75\textwidth]{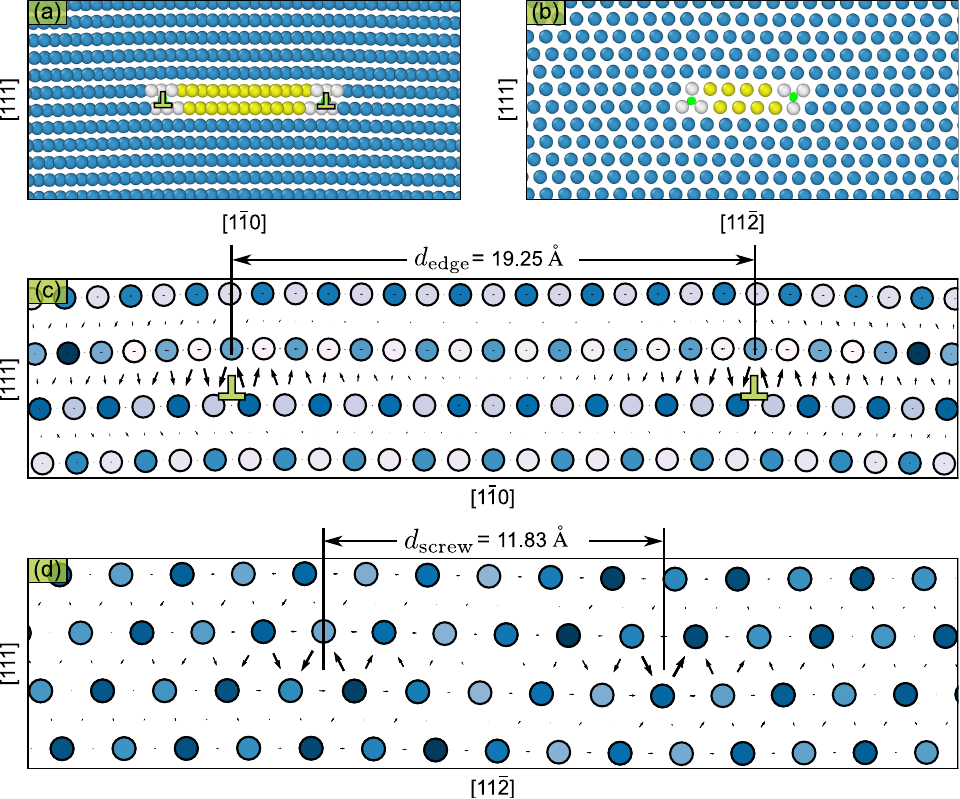}
\caption{\label{fig:disl_mob}Dislocation core structures of two Shockley partials of \adisl edge and screw dislocations in FCC Ni predicted by DP-Ni. (a) and (b) show the atomic configurations of edge and screw dislocations visualized by OVITO \cite{stukowski2009visualization}. The common neighbor analysis (CNA) method~\cite{honeycutt1987molecular} is utilized to distinguish the FCC (blue), HCP (yellow), and other (white) local atom stacking. The dislocation extraction algorithm (DXA)~\cite{Stukowski_2010} is employed to identify dislocations precisely. (c) and (d) show the corresponding differential displacement plots for (a) and (b). The partial core separations are shown to be $d_{\mathrm{edge}}$ and $d_{\mathrm{screw}}$ of 19.25 {\AA} and 11.83 \AA.}
\end{figure*}

The Shockley partial dislocations are separated by a stable stacking fault~\cite{anderson2017theory}. 
Accurate modeling of dislocation dissociation and partial dislocation separation is essential for precise modeling of plastic behavior. 
We simulate this dissociation by inserting a perfect \adisl edge and screw dislocation at the center of a $301 \times 17 \times 85$ {\AA}$^3$ (the dislocation line is along the $y$-direction, while the Burgers vector is in the $x$-direction) and $15 \times 302 \times 85$ {\AA}$^3$ (dislocation line and Burgers vector parallel the $x$ direction) supercells with periodic boundary conditions in the $x$- and $y$-directions, respectively. 
We then minimize the energy (molecular statics simulation at 0 K) with DP-Ni; the relaxed configurations are shown in Fig.~\ref{fig:disl_mob}. The edge and screw configurations decompose into a pair of Shockley partial dislocations with different separation distances. Differential displacement (DD)~\cite{vitek1970core} plots reveal the strain fields around a dislocation by measuring the relative displacement of a pair of nearest neighbor atoms. 
A partial dislocation consists of three atoms with clockwise or counterclockwise net chirality.  
In this case, the DD plots in Figs.~\ref{fig:disl_mob}(c) and (d) identify the positions of the partial dislocations. 
The DP-Ni partial dislocation separation distances are $d_{\mathrm{edge}}=19.25$ {\AA} and $d_{\mathrm{screw}}=11.83$ {\AA}. 
Our result for the edge dislocation separation distance aligns with weak-beam transmission electron microscopy observations, i.e., 26$\pm8$ {\AA}~\cite{carter1977stacking}. 
While $d_{\mathrm{screw}}$ is not easily measured experimentally, our result of 11.83 {\AA} is consistent with the 12.0 {\AA} obtained from the previous DFT calculation~\cite{tan2019dislocation}.

\subsection{Structures and Energies of Tilt Grain Boundaries}

Grain boundaries (GB) in polycrystalline materials limit dislocation slip and, hence, play an important role in determining strength and ductility. 
In this study, we investigate several high angle symmetric tilt GBs, constructed based upon geometry.
We then identify the lowest energy GB structure by sliding one grain relative to the other and minimizing the energy. 
The lowest-energy GB configurations (after relaxation using DP-Ni) are shown in Fig.~S2. The relaxed $\Sigma$3 $[1\bar{1}0]$ $(111)$, $\Sigma$5 $[100]$ $(0\bar{2}1)$ and $\Sigma$11 $[1\bar{1}0]$ $(113)$ remain symmetric, while $\Sigma$3 $[1\bar{1}0]$ $(112)$, $\Sigma$7 $[111]$ $(3\bar{2} \bar{1})$ and $\Sigma$9 $[1\bar{1}0]$ $(22\bar{1})$ relax to an asymmetric boundary structure. 
Table~\ref{tab:surface and defects energies} shows the GB energies from both DFT calculations and with several interatomic potentials. 
DP-Ni accurately reproduces all GB energies with only minor discrepancies ($<6.7$\%) compared to the respective DFT values. 
Both DP-Ni and DFT identify $\Sigma$3 $[1\bar{1}0]$ $(111)$ as the lowest GB energy, as reported in most experimental observations~\cite{RANDLE20061777}. 
The energy ordering follows the pattern: $\Sigma$3 $(111)$ $<$ $\Sigma$11 $<$ $\Sigma$3 $(112)$ $<$ $\Sigma$9 $<$ $\Sigma$7 $<$ $\Sigma$5. 
Other potentials capture the energy ordering of these GBs but are less quantitative relative to the DFT results. 
EAM accurately predicts the energy of $\Sigma$3 $[1\bar{1}0]$ $(111)$, but overestimates other GB energies by 16.9\%-22.1\%. MEAM\_2021, MEAM\_2015 and qSNAP roughly reproduce the energy of low $\Sigma$ GBs but drastically underestimate the energy of the important $\Sigma$3 $[1\bar{1}0]$ $(111)$ (by $>50$\%) - in fact the MEAM\_2021 gives an unphysical negative value for this GB energy. 
Overall, DP-Ni reproduces all GB structures and corresponding energies, demonstrating its potential for simulation of GB behavior (e.g., GB migration, deformation twinning, and disconnection behavior~\cite{zhu2019situ,KHATER20122007,lu2015transition}).

\subsection{Allotropic Transformation of Nickel under Uniaxial Tension}

\begin{figure*}[!htbp]
  \includegraphics[width=0.63\textwidth]{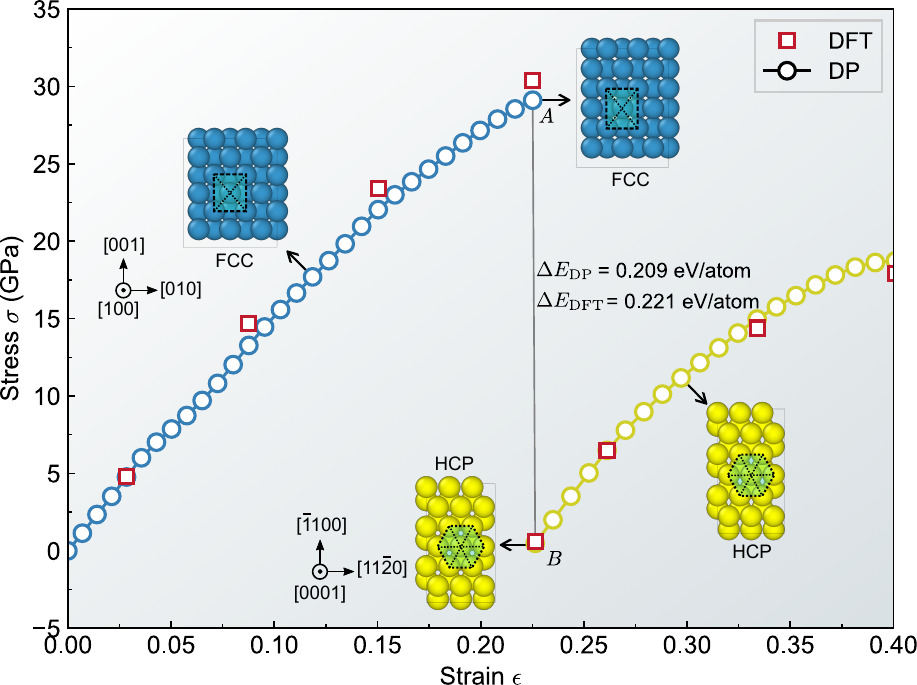}
  \caption{\label{fig:uniaxial_tension} The uniaxial stress vs. strain of Ni under uniaxial tension along $[001]$ using DP-Ni at 0 K. The red square points are the stress calculated by DFT under the corresponding strain. Allotropic phase transformation is induced upon a precipitous decrease in stress. The energy difference values represent the cumulative energy discrepancy of structures at points A and B as computed by the DP-Ni and DFT, respectively. The insert atomic configurations are labeled using CNA~\cite{honeycutt1987molecular} for FCC (blue) and HCP (yellow) local packing by OVITO~\cite{stukowski2009visualization}. The crystallographic orientation relationship for FCC-HCP is \{100\}\textsubscript{FCC}$\parallel$\{0001\}\textsubscript{HCP} and $\left\langle 010 \right\rangle$\textsubscript{FCC}$\parallel\left\langle 11\bar{2}0 \right\rangle$\textsubscript{HCP}.}
\end{figure*}

As a further, stringent test, on the performance of  DP-Ni, we examine the allotropic phase transformation of nickel under uniaxial loading. 
Figure~\ref{fig:uniaxial_tension} shows the stress-strain relationship for Ni under uniaxial tension along the $[001]$ crystallographic direction; calculations are conducted at 0 K. 
The results show a monotonic increase in stress as a function of strain, followed by a sudden drop at a strain of $\sim$0.226. 
The inset atomic configuration depicts the observed atomic structure at different strains, showing the strained FCC and HCP structures. 
These insets indicate that the abrupt drop in the stress corresponds to an FCC $\rightarrow$ HCP transformation. 
Subsequently, the HCP phase remains stable for an additional strain of at least 15\% (from point B). 
The transformation strain is quite large compared with other strain-induced transformations~\cite{CHEN201965,GUAN2023103707,LI2023118854,YANG2017347,HIRTH19992409}. 
However, such large transformation strains are not unusual for FCC metals; e.g., see the experimental observations and theoretical calculations~\cite{PhysRevLett.128.015701,XIE2015191,WEI202036,ZHANG20085451,PhysRevLett.123.205701,LI2022117850,diao2003surface}. 

Specifically, in the case of nickel, an FCC$\rightarrow$HCP transformation was observed experimentally in nanocrystalline  (nanoscale grained) Ni subjected to large plastic strains~\cite{luo2019plastic,guo2021plastic}. 
To confirm this transformation, the energy differences ($\Delta E$) between structures at points A and B are measured, as shown in Fig.~\ref{fig:uniaxial_tension}. 
The positive $\Delta E$ obtained from both the DP-Ni and DFT calculations suggest that the HCP structure at point B is more stable compared to the strained FCC structures at point A. 
We apply DFT to calculate the stress of the structures in the strain-stress curve from DP-Ni. 
As indicated by the red squares in Fig.~\ref{fig:uniaxial_tension}, the results from DP-Ni are very close to those from DFT. 
At the allotropic transformation strain, the energy difference between $\Delta E_{\mathrm{DP}}$ and $\Delta E_{\mathrm{DFT}}$, is very small ($\sim 10$ meV/atom), corroborating the fidelity of the DP-Ni model. 
The observed crystallographic orientation relationship between FCC and HCP structures in our study presents an atypical case, namely, \{100\}\textsubscript{FCC}$\parallel$\{0001\}\textsubscript{HCP} and $\left\langle 010 \right\rangle$\textsubscript{FCC}$\parallel\left\langle 11\bar{2}0 \right\rangle$\textsubscript{HCP}. 
This orientation deviates from the commonly documented strain-induced FCC$\rightarrow$HCP transformation, which is typified by the orientation relation \{111\}\textsubscript{FCC}$\parallel$\{0001\}\textsubscript{HCP} and $\left\langle1\bar{1}0 \right\rangle$\textsubscript{FCC}$\parallel\left\langle11\bar{2}0\right\rangle$\textsubscript{HCP}~\cite{pattamatta2022allotropy,PhysRevLett.128.015701,WU2021106026,YU20179}. 
The present orientation relationship is associated with the very large mechanical strains here. 
This orientation relationship was previously reported based upon theoretical~\cite{XIE2015191,PhysRevB.44.9155} and experimental investigations~\cite{cryst11101164} in other FCC metals. 
Interestingly, another unconventional orientation relation was observed in nanocrystalline nickel $\left\langle110\right\rangle$\textsubscript{FCC}$\parallel\left\langle1\bar{2}1\bar{3} \right\rangle$\textsubscript{HCP}~\cite{luo2019plastic}.

\section{Conclusion}
We developed a ``magnetism-hidden'' machine learning Deep Potential (DP) model for both FCC and HCP nickel, based upon DFT calculations. 
The nickel DP (DP-Ni) was trained using spin-polarized DFT calculations employing a relatively small training dataset  (see Supplementary Table S2). 
Inclusion of spin polarization was found to be essential.
DP-Ni achieves DFT-level accuracy in predicting a wide range of properties for both FCC and HCP Ni, such as (finite-temperature) lattice parameters and elastic constants, phonon spectra, cohesive and decohesion energies/stresses, point defect formation energies, stacking fault energies, and dislocation and grain boundary properties. 
The DP-Ni results are, overall, more reliable than predictions based upon other potentials (including semi-empirical and other machine learning potentials).
DP-Ni thus serves as a promising tool for large-scale atomistic simulations of Ni, especially for mechanical properties. 
Our DP-Ni model facilitated the examination of the allotropic FCC$\rightarrow$HCP phase transition, wherein we identified a high critical strain and an atypical orientation relationship under uniaxial tensile loading.  
The new DP-Ni potential and the associated training datasets can be utilized as a foundation for developing ML potential for Ni-based superalloys, medium-entropy (FeCoNi) and high-entropy (FeCoNi-based) alloys through methods such as the DP attention pre-training model~\cite{zhang_2022_arxiv}.

\section{METHODS}

\subsection{DFT Calculations}
The Vienna Ab initio Simulation Package (VASP)~\cite{kresse1996efficiency,kresse1996efficient} is used to perform the density functional theory (DFT) calculations using the projector augmented wave (PAW) method~\cite{blochl1994projector} for generation of the training set and determining property benchmarks. The exchange-correlation function is treated within the generalized gradient approximation (GGA), as formulated by Perdew-Burke-Ernzerhof (PBE)~\cite{perdew1996generalized}. 
The basis set includes Ni 3d\textsuperscript{8}4s\textsuperscript{2} electron levels. 
We employ a plane wave cutoff energy of 600 eV and the Methfessel-Paxton method~\cite{methfessel1989high} to determine partial wave function occupations with a 0.12 eV smearing width. 
Monkhorst-Pack k-point grids~\cite{monkhorst1976special} are optimized to sample the Brillouin zone with a  0.1 {\AA}\textsuperscript{-1} k-points grid. 
A 10\textsuperscript{-6} eV/atom total energy and a 10\textsuperscript{-3} eV/\AA~ ionic force convergence criteria is employed. 
Both the ground state calculations and \textit{ab initio} molecular dynamics (AIMD) simulations account for spin-polarization (magnetic moment).
More details may be found in the Supplementary Information (SI).

\subsection{Molecular Dynamics Simulations}
Molecular dynamics (MD) and static calculations are conducted using the Large-scale Atomic/Molecular Massively Parallel Simulator (LAMMPS)~\cite{thompson2022lammps}. Atomic structure optimization is performed using the conjugate gradient method; convergence criteria for force is 10\textsuperscript{-10} eV/{\AA} (self-interstitial configurations are converged to energy 10\textsuperscript{-13}).  
The same simulation cell size/configurations are employed in both DFT and MD calculations of the elastic constants, surface energy, point-defect formation energy, grain boundary, stacking fault energy, cohesive and decohesive energies, phonon spectra, and ideal strength. See the SI for more details.

\subsection{Training Strategy of Deep Potential for Ni}
\label{sec:Training Strategy}

We utilize the general Deep Potential Generator (DP-GEN) scheme~\cite{zhang2020dp}, the Deep Potential Smooth Edition (DeepPot-SE)~\cite{zhang2018end}, along with the ``specialization'' strategy~\cite{wen2021specialising} to generate the training datasets (a 6 {\AA} cutoff radius is used throughout). 
We employ a neural network of 240 $\times$ 240 $\times$ 240. 

Initially, supercells with three perfect 2$\times$2$\times$2 cell BCC, FCC, and HCP (2, 4, and 2 atoms per cell) are constructed. 
Supercell volumes are rescaled by a scaling factor (0.96-1.06 in steps of 0.02), resulting in six configurations for each phase. 
These scaled supercells are then randomly perturbed (3X) by scaling the supercell translation vectors and adding relative atomic translation in the range of -3 to 3\% and  -0.01 to 0.01 Å, respectively. 
Next, two steps of AIMD are conducted for each distorted structure (at 100 K) in the NVT ensemble (Nosé-Hoover thermostat). 
A total of 108 ionic configurations are obtained from the AIMD calculations (converged electronic degrees of freedom), providing atom coordinates, total energy, atomic forces, and virial tensors. 
This data serves as the initial training dataset for the DP-GEN loop. 

In each DP-GEN  training step, four DP models are initiated using four random initial neural net parameter sets. 
The training step consists of 400,000 epochs.
The learning rate starts at 10$^{-3}$ and exponentially decays to 5 $\times$ 10\textsuperscript{-8} during the training. 
The loss function prefactors for the energy, atomic force, and virial tensor $p_\mathrm{e}^{\mathrm{start}}$ = 0.02, $p_\mathrm{e}^{\mathrm{limit}}$ = 2, $p_\mathrm{f}^{\mathrm{start}}$ = 1,000, $p_\mathrm{f}^{\mathrm{limit}}$ = 1, $p_\mathrm{v}^{\mathrm{start}}$ = 0, and $p_\mathrm{v}^{\mathrm{limit}}$ = 0, respectively, vary during training.

During the DP-GEN loop exploration step, a single DP model is selected to explore various bulk and surface structures for each of the distorted BCC, FCC, and HCP supercells using DP-based MD (DPMD) with the LAMMPS package. 
The bulk structure is explored via MD in the temperature range of 50 to 3,283.2 K (1.9 times the Ni melting point $T_\mathrm{melt}$) under isothermal-isobaric (NPT) conditions, with pressures varying between 0.001 and 50 kBar. 
Surface structures are constructed from all crystal supercells by introducing $\{100\}$, $\{110\}$, and $\{111\}$ (BCC and FCC) and $\{0001\}$ and $\{10\bar{1}0\}$ (HCP) surfaces. 
Surface supercells are scaled and perturbed similarly to the bulk structures and simulated via DPMD in a canonical (NVT) ensemble over the same temperature range. 
A criterion is set for choosing amongst the four models at each DPMD step to perform spin-polarized DFT calculations (energy, force, virial) to add to the training datasets for subsequent DP-GEN loop iterations.
See the SI for more details.

While the final four DP models reproduce many properties of FCC and HCP Ni, they do not accurately reproduce cohesive properties (see Table S1 and Fig. S3(b)). 
We address this by generating a specialized training dataset consisting of 170 configurations specifically selected from the cohesive energy line. 
These configurations include 17 distinct structures; each assigned a weight of 10 in the final training set (i.e., 10X the other structures). 
The final training is performed on both the training datasets from DP-GEN and the ``specialization'' (DFT calculations are all spin-polarized). 
More details are provided in the SI (see Table S2 for a summary of training datasets employed).
We emphasize that while our training set is large, it is considerably smaller than those employed in other ML potentials~\cite{wen2021specialising,byggmastar2020gaussian,smith2021automated}.

\section{DATA AVAILABILITY}
The DP-Ni model and training datasets will be made available upon acceptance of the paper.

\section{ACKNOWLEDGMENTS}

This work is supported by the Research Grants Council, Hong Kong SAR through the General Research Fund (17210723). TW acknowledges additional support by The University of Hong Kong (HKU) via seed funds (2201100392, 2309100163). Part of the computational resources are provided by HKU research computing facilities.

\def\bibsection{\section*{\refname}} 
\bibliography{./Ni_DP_apssamp_comb.bib}

\newpage 

\section*{Supplementary Information}
\beginsupplement
\subsection{Property calculation by DFT}
    The stress-strain method~\cite{shang2007first,GONG2020109174} is applied to evaluate the elastic constants of FCC and HCP Ni. 
The final values of $C_{ij}$ are obtained by averaging the data after applying a set of normal strains (-0.01, -0.005, 0.005, 0.01) and shear strains (-0.01, -0.005, 0.005, 0.01) to the FCC and HCP unit cells (4 and 2 atoms, respectively). 
The monovacancy and self-interstitials formation energies in FCC Ni are estimated using a 3 $\times$ 3 $\times$ 3 supercell with full relaxation. 
Generalized stacking fault $\gamma$-lines for the FCC $\{111\}$ planes are calculated using the slab-vacuum supercell method~\cite{vitek1968intrinsic}, where atoms are allowed to relax only in the direction perpendicular to the slip plane. 
These supercells consist of 24 atomic layers and feature a vacuum layer thickness of 20 {\AA}. 
The translation plane is selected at the center of the supercell. 
Surface energy and grain boundary energy calculations are performed using configurations with a 20 {\AA} vacuum layer. 
For the $\Sigma$7 $[111]$ $(3\bar{2} \bar{1})$ configuration, a k-points grid spacing of 0.15 {\AA}\textsuperscript{-1} is utilized, while other calculation parameters are consistent with the DFT settings in the main text.

    The ideal strength of FCC Ni at 0 K is calculated using DFT through the method of incremental loading. 
Tensile and pure shear strengths along high symmetry crystallographic directions are identified as the maxima of stress along the incremental loading path, simulated by subjecting a suitably oriented periodic supercell of FCC Ni to combined stress/strain loading, so as to maintain the desired state of stress in the material. 
For tensile strength determination along a crystallographic direction, at each step of incremental loading, the strain along the tensile axis is held constant (fixed cell vector in that direction), while relaxing all stress components (except along the straining direction) to zero; i.e.,  a uniaxial stress state. 
For the FCC crystal, tensile strengths are determined along the high symmetry directions $[001]$, $[011]$ and $[111]$. 
Shear strength along a direction on a crystallographic plane is determined by applying a shear strain along that direction and relaxing all other stress components; i.e., a pure shear stress state. 
Shear strengths are determined on the $(111)$ plane along the symmetric $[\bar{1}10]$ and $[1\bar{1}0]$ directions, and along the asymmetric $[11\bar{2}]$ (easy) and $[\bar{1}\bar{1}2]$ (hard) directions. 
The ideal strength under hydrostatic dilatation (i.e., tensibility), is defined as the maximum hydrostatic stress that the crystal can sustain and is determined by incrementally dilating (isotropically expanding) the FCC crystal. 
At each volume along the loading path, the resultant hydrostatic stress (negative pressure) is evaluated. 
The maximum hydrostatic stress along the loading path is the tensibility. 
Constrained relaxations under combined stress and strain boundary conditions for the ideal strength calculations are achieved through an in-house developed patch to the standard VASP code (see the Supplementary Information in~\cite{pattamatta2022allotropy}). 
In this algorithm, the cell vectors and cell shape, barring those fixed by the applied strain and the ionic positions are iteratively optimized until the stress boundary conditions are satisfied.

\subsection{Property calculation by MD}
    Phonon spectra calculations are performed using the PHONOPY~\cite{togo2015first} and phono-LAMMPS~\cite{phonoLAMMPS} software packages. A 3 $\times$ 3 $\times$ 3 supercell is employed for the FCC structure (108 atoms), while a 4 $\times$ 4 $\times$ 4 supercell is employed for the HCP structure (128 atoms).

The lattice parameters and elastic constants at finite temperatures are determined using a time-averaging approach on individual properties, employing the Nosé-Hoover thermostat in LAMMPS. 
A fully periodic 16 nm $\times$ 16 nm $\times$ 16 nm supercell is initially constructed for a perfect crystal. 
This supercell is then equilibrated for 40,000 time steps (40,000 fs) under stress-free conditions and at the respective temperatures using MD simulations in an NPT ensemble. 
After equilibration, the simulation box size is determined by time averaging over 4,000 fs using the same NPT ensemble. 
The lattice parameter is determined from the size of the simulation box, averaging over 10 measurements. 
Elastic constants are also obtained through a time-averaging scheme in MD. 
The same supercell is subjected to equilibration for 168,000 time steps under stress-free conditions at the target temperatures in an NPT ensemble. 
An additional 168,000 time steps are performed under the canonical NVT ensemble, applying a $\pm1$\% strain for each strain component. 
The stress is measured by averaging the instantaneous stress values obtained at alternate time steps over 14,000 time steps, under the same NVT ensemble. 
Each elastic constants are determined from the resulting stresses at each strain (168,000/14,000=12 measurements for each strain symmetry). 
For all interatomic potentials, the simulations are repeated three times (three random seeds for the initial atomic velocity distributions) and the average elastic constants are determined at each temperature.

    In this study, we examine the $a/2$ $\langle 110 \rangle$ edge and screw dislocation cores located on the $\{111\}$ plane using a periodic array configuration of dislocations. 
The slip plane is the $x$-$y$ plane. 
Periodic boundary conditions are implemented in the $x$ and $y$ directions, while the top and bottom ($z$) surfaces are treated as traction-controlled/free surfaces. 
For the pure edge and screw dislocations, the Burgers vector \textbf{b} is aligned with the $x$ direction. 
To minimize the interaction between adjacent dislocation cores, we create sufficiently large supercells; 301 $\times$ 17 $\times$ 85 {\AA$^3$}  for edge dislocations and 15 $\times$ 302 $\times$ 85 {\AA$^3$}  for screw dislocations. 
For edge dislocations, the dislocation line is along the $y$-direction with the Burgers vector parallels the $x$-direction. 
For screw dislocations, the dislocation line and Burgers vector are parallel to the $x$-direction. 
To introduce the initial full dislocation, we apply the displacement field of the corresponding Volterra dislocation at the center of the supercells using the Atomsk package~\cite{hirel2015atomsk}. 
Dislocations with a nonzero screw component undergo a homogeneous shear strain of $\epsilon_{yx}$  = \textbf{b} · $\bm{\xi}$/2 to rectify the plastic shear strain induced by the screw component. 
The constructed dislocation cores are subsequently relaxed using the conjugate gradient method, with a force convergence criterion set to 10\textsuperscript{-4} eV/Å. 
The final atomic configurations are visualized using the open visualization tool (OVITO)~\cite{stukowski2009visualization}. 
Additionally, differential displacement plots are analyzed utilizing the atomistic manipulation toolkit Atomman~\cite{atomman}.

\subsection{Training Strategy of Deep Potential for Ni}

Figure S3(a) illustrates the workflow used for developing the DP model for Ni. In the training step of the DP-GEN loop, distorted 2$\times$2$\times$2 supercells of BCC, FCC and HCP structures are employed as the starting configurations within the DPMD calculations. To enhance the sampling efficiency, several temperatures are explored for the bulk structures, partitioned into four regions:  (a) 50 K, [0.1, 0.2, 0.3, 0.4]$T_\mathrm{melt}$; (b) [0.5, 0.6, 0.7, 0.8, 0.9]$T_\mathrm{melt}$; (c) [1.0, 1.1, 1.2, 1.3, 1.4]$T_\mathrm{melt}$ and (d) [1.5, 1.6, 1.7, 1.8, 1.9]$T_\mathrm{melt}$. 
In each temperature region, the pressure is systematically varied [0.001, 0.01, 0.1, 1, 5, 10, 20, 50] kBar. 
A total of 40 different MD conditions (temperatures and pressures) are generated within each temperature region. 
Surface structure exploration begins once all bulk structures have been explored. 
Low Miller index surfaces, i.e.,$\{100\}$, $\{110\}$, and $\{111\}$ planes for BCC and FCC, $\{0001\}$ and $\{10\bar{1}0\}$ surfaces for HCP structures are constructed. 
Surface supercells are scaled and perturbed similar to the process used for bulk structures. 
The canonical (NVT) ensemble is employed within the same temperature range as bulk structures for DPMD simulations. 
After the MD simulation is complete, an indicator $\varepsilon$ measures the standard deviations of atomic force predictions among the four DP models to select candidate configurations based upon trust levels $\sigma_{lo}=0.10$ and $\sigma_{hi}=0.25$ across the entire temperature range for both bulk and surface structures. 
Configurations within $\sigma_{lo} < \varepsilon < \sigma_{hi}$ are chosen as candidates for spin-polarized DFT calculations. 
The resulting DFT data is then added to the training datasets and used to generate four new DP models for the subsequent DP-GEN loop. 
Convergence is achieved when the number of candidate structures for the DFT calculations is $<0.1$\% of the total number of configurations explored.

Based on the DP-GEN loop, the final four DP models demonstrate the capability to reproduce the general properties of Ni such as total energies and elastic constants (for example, see DP\_nonspecX in Table S1). 
However, the cohesive energy curve from DP is inconsistent with DFT results (see Fig. S3(b)). We address this issue by generating special training datasets comprising 17 distinct configurations specifically selected from the cohesive energy curve and each assigned a weight of 10 (10X the structures from DP-GEN loop). 
Virial tensors and forces are not considered for the specialization dataset. 
A final training is performed on both the training datasets (2,020 entries in Table S2) from DP-GEN and ``specialization''. The learning rate starts at 0.001 and decays exponentially to 5 $\times$ 10\textsuperscript{-8}. 
The training consists of 8,000,000 epochs and the pre-factors in the loss function are $p_\mathrm{e}^{\mathrm{start}}$ = 0.02, $p_\mathrm{e}^{\mathrm{limit}}$ = 2, $p_\mathrm{f}^{\mathrm{start}}$ = 1000, $p_\mathrm{f}^{\mathrm{limit}}$ = 1, $p_\mathrm{v}^{\mathrm{start}}$ = 0.02, and $p_\mathrm{v}^{\mathrm{limit}}$ = 1, respectively.

\begin{table*}[!htbp]
      \caption{\label{tab:basic_properties_large}
      Lattice parameters ($a$, $c/a$), bulk energies ($E$), cohesive energies ($E_\mathrm{coh}$), and elastic constants ($C_{ij}$) of FCC and HCP obtained by using different interatomic potentials for Ni. DFT\_FM and DFT\_NM represent DFT calculations that are spin-polarized (magnetic) and non-spin polarized (non-magnetic). 
The DP\_nonspecX is a DP model before ``specialization''. 
Bold numbers indicate deviations of more than 15\% compared to DFT\_FM and/or experiment.
      }
      \vspace{2.2mm} 
      \begin{ruledtabular}
      \begin{tabular}{cccccccccccccccc}
         & \multicolumn{1}{c}{FCC} &&&&&& \multicolumn{1}{c}{HCP}\\ 
          \cline{2-7}\cline{8-16}
          \vspace{1.2mm}
          
      \textrm{Potential}&
      \textrm{$a$}&
      \textrm{$E$}&
      \textrm{$E_\mathrm{coh}$}&
      \textrm{$C_{11}$}&
      \textrm{$C_{12}$}&
      \textrm{$C_{44}$}&
      \textrm{$a$}&
      \textrm{$c/a$}&
      \textrm{$E$}&
      \textrm{$E_\mathrm{coh}$}&
      \textrm{$C_{11}$}&
      \textrm{$C_{12}$}&
      \textrm{$C_{13}$}&
      \textrm{$C_{33}$}&
  
      \textrm{$C_{44}$}\\
      \colrule
                  
        Expt. &3.520\footnote{\cite{kanhe2016investigation}, $^\text{b}$\cite{kittel2005introduction}, $^\text{c}$\cite{simmons1971single}, $^\text{d}$\cite{lagrow2013can}} &-      &4.450$^\text{b}$ &261.2$^\text{c}$&150.8$^\text{c}$&131.7$^\text{c}$ &2.487$^\text{d}$&1.645$^\text{d}$&- &- &- &- &- &- &-\\
      DFT\_FM   &3.517 &-5.467 &4.865 &275.7 &156.0 &130.7 &2.484 &1.643 &-5.443 &4.841 &312.0 &142.3 &122.8 &330.7 &55.5\\
      DFT\_NM   &3.511 &-5.411 &5.186    &256.7 &174.3 &114.7 &2.475 &1.652 &-5.386 &5.160 &296.0 &\textbf{173.0} &128.8 &337.2 &49.3\\

    \rule{0pt}{3.2mm}
    \textbf{\textit{EAM}}\\  
    \rule{0pt}{3.2mm}
    Mishin\cite{mishin1999interatomic} &3.520 &-4.450 &4.450 &247.9 &147.8 &124.8 &2.483 &1.619 &-4.430 &4.430 &302.2 &147.6 &\textbf{76.9} &\textbf{213.3} &\textbf{64.3}\\
    Zhou\cite{zhou2004misfit} &3.520 &-4.450 &4.450 &247.0 &147.3 &124.9 &2.483 &1.658 &-4.434 &4.434 &302.6 &138.1 &\textbf{77.3} &\textbf{247.7} &55.6\\
    Foiles\cite{foiles1986embedded} &3.520 &-4.450 &4.450 &233.3 &154.3 &127.6 &2.489 &1.630 &-4.448 &4.448 &295.7 &150.0 &\textbf{95.9} &349.6 &\textbf{69.4}\\
    Ackland\cite{ackland1987simple} &3.524 &-4.459 &4.459 &260.7 &150.5 &131.4 &2.490 &1.642 &-4.455 &4.455 &325.0 &141.9 &\textbf{96.9} &318.9 &\textbf{77.2}\\
    Adams\cite{adams1989self} &3.520 &-4.450 &4.450 &235.6 &153.1 &133.5 & 2.489 &1.630 &-4.447 &4.447 &300.9 &148.7 &\textbf{91.9} &357.3 &\textbf{72.5}\\
    Mendelev\cite{mendelev2012development} &3.518 &-4.390 &4.390 &247.0 &147.3 &122.8 &2.486 &1.658 &-4.353 &4.353 &267.0 &134.2 &\textbf{58.0} &\textbf{196.7} &\textbf{39.0}\\
    Stoller\cite{stoller2016impact} &3.520 &-4.450 &4.450 &240.9 &150.5 &127.1 &2.482 &1.654 &-4.428 &4.428 &298.5 &145.2 &\textbf{71.6} &\textbf{260.5} &48.6\\
    Angelo\cite{angelo1995trapping} &3.520 &-4.450 &4.450 &246.7 &147.5 &125.0 &2.480 &1.685 &-4.437 & 4.437 &337.1 &\textbf{167.1} &109.9 &359.7 &58.9\\

    \rule{0pt}{3.2mm}
    \textbf{\textit{MEAM}}\\ 
    \rule{0pt}{3.2mm}
    Etesami\cite{etesami2018molecular} &3.520 &-4.450 &4.450 &255.3 &155.8 &129.8 &2.486 &1.640 &-4.441 &4.441 &315.8 &155.1 &\textbf{97.0} &366.2 &59.7\\
    Vita\cite{vita2021exploring} &3.519 &-3.952 &3.952 &278.3 &169.8 &112.5 &2.490 &1.630 &-3.956 &3.956 &327.6 &159.5 &131.9 &355.6 &\textbf{73.0}\\
    Asadi\cite{asadi2015two} &3.521 &-4.450 &4.450 &260.4 &150.7 &131.0 &2.487 &1.641 &-4.439 &4.439 &326.8 &139.3 &\textbf{95.9} &364.8 &\textbf{74.6}\\
    Aitken\cite{aitken2021modified} &3.504 &-4.849 &4.849 &264.2 &149.6 &125.6 &2.471 &1.653 &-4.826 &4.826 &331.4 &134.7 &\textbf{94.4} &358.2 &\textbf{85.4}\\
    Shim\cite{shim2003modified} &3.521 &-4.450 &4.450 &261.2 &150.8 &131.7 &2.484 &1.647 &-4.429 &4.429 &332.7 &139.0 &\textbf{91.9} &366.5 &\textbf{75.0}\\
    Shim\cite{shim2013prediction} &3.521 &-4.450 &4.450 &261.2 &150.8 &131.7 &2.484 &1.647 &-4.429 &4.429 &332.7 &139.0 &\textbf{91.9} &366.5 &\textbf{75.0}\\
    Ko\cite{ko2015development} &3.521 &-4.450 &4.450 &260.4 &148.6 &\textbf{111.1} &2.487 &1.642 &-4.440 &4.440 &314.7 &133.8 &108.3 &336.0 &\textbf{77.2}\\
    Maisel\cite{maisel2017thermomechanical} &3.521 &-4.450 &4.450 &260.4 &148.6 &\textbf{111.1} &2.487 &1.642 &-4.440 &4.440 &314.7 &133.8 &108.3 &336.0 &\textbf{77.2}\\
    Lee\cite{lee2003semiempirical} &3.521 &-4.450 &4.450 &261.2 &150.8 &131.7 &2.484 &1.647 &-4.429 &4.429 &332.7 &139.0 &\textbf{91.9} &366.5 &\textbf{75.0}\\
    Mahata\cite{mahata2022modified} &3.521 &-4.450 &4.450 &260.4 &150.7 &131.0 &2.487 &1.641 &-4.439 &4.439 &326.7 &139.3 &\textbf{95.9} &364.7 &\textbf{74.6}\\

    \rule{0pt}{3.2mm}
    \textbf{\textit{ML(SNAP)}}\\ 
    \rule{0pt}{3.2mm}
    Zuo\cite{zuo2020performance} &3.521 &-5.780 &\textbf{5.780} &267.5 &155.3 &125.7 &2.491 &1.643 &-5.772 &\textbf{5.772} &334.0 &144.0 &109.1 &369.2 &\textbf{77.2} \\
    Zuo\cite{zuo2020performance} &3.522 &-5.781 &\textbf{5.781} &282.9 &168.3 &129.3 &2.492 &1.650 &-5.773 &\textbf{5.773} &341.8 &\textbf{166.9} &132.5 &\textbf{421.0} &63.6\\
    \textbf{\textit{DP}}\\ 
    DP\_nonspecX &3.517 &-5.467 &\textbf{2.978} &278.4 &157.3 &129.1 &2.484 &1.643 &-5.446 &\textbf{2.958} &322.5 &150.0 &113.4 &356.1 &61.1\\

    \end{tabular}
    \end{ruledtabular}
    \end{table*}

\begin{table*}[!htbp]
  \caption{\label{tab: dp_training_data}
    Summary of the training datasets for DP-Ni.}
  \centering
  \begin{ruledtabular}
    \begin{tabular}{ccc}
      \textrm{Dataset type}&
                             \textrm{Number of datasets}&
                                                          \textrm{Weightage}\\
      \colrule
      
      Initialization datasets  &108   &1\\
      DP-GEN bulk & 1140 & 1 \\
      DP-GEN surface & 602 & 1 \\
      Cohesive energy (specialization) & 17 & 10 \\
      Total\footnote{The cohesive energy datasets are considered as 17 $\times$ 10 = 170.} & 2020 &  \\      
    \end{tabular}
  \end{ruledtabular}
\end{table*} 

\begin{figure*}[!htbp]
  \includegraphics[width=0.85\textwidth]{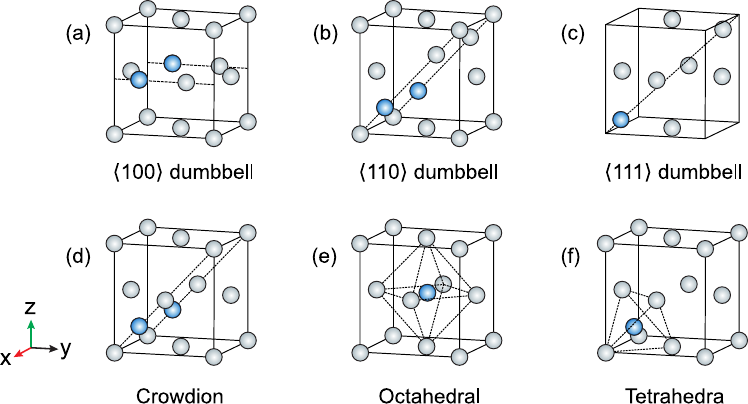}
  \caption{\label{fig:interstitial}Six types of self-interstitial configurations in the FCC crystal lattice.
  }
\end{figure*}

\begin{figure*}[!htbp]
  \includegraphics[width=0.9\textwidth]{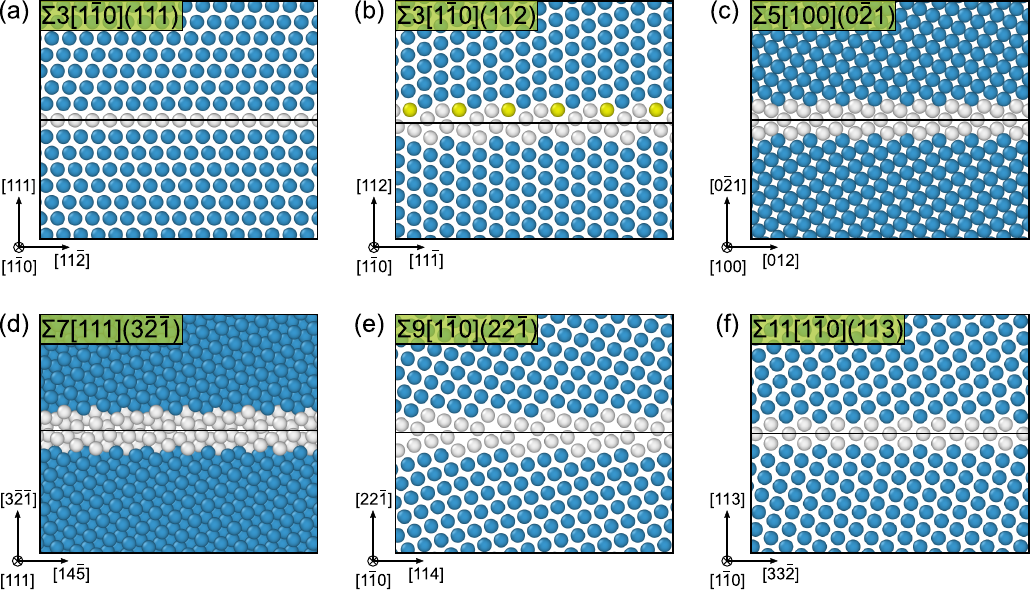}
  \caption{\label{fig:gb} Optimized tilt grain boundary structures in FCC Ni obtained using DP-Ni at zero temperature. 
    The relaxed (a) $\Sigma$3 $[1\bar{1}0]$ $(111)$, (c) $\Sigma$5 $[100]$ $(0\bar{2}1)$ and (f) $\Sigma$11 $[1\bar{1}0]$ $(113)$ configurations are symmetric tilt grain boundaries and  (b) $\Sigma$3 $[1\bar{1}0]$ $(112)$, (d) $\Sigma$7 $[111]$ $(3\bar{2} \bar{1})$ and (e) $\Sigma$9 $[1\bar{1}0]$ $(22\bar{1})$ are asymmetric boundaries. 
    The atomic configurations are visualized using the open visualization tool (OVITO)~\cite{stukowski2009visualization} with colors based upon the common neighbor analysis (CNA) method~\cite{honeycutt1987molecular}. 
    The FCC, HCP and other types of atoms are marked in blue, yellow and white, respectively.}
\end{figure*}

\begin{figure*}[!htbp]
  \includegraphics[width=0.9\textwidth]{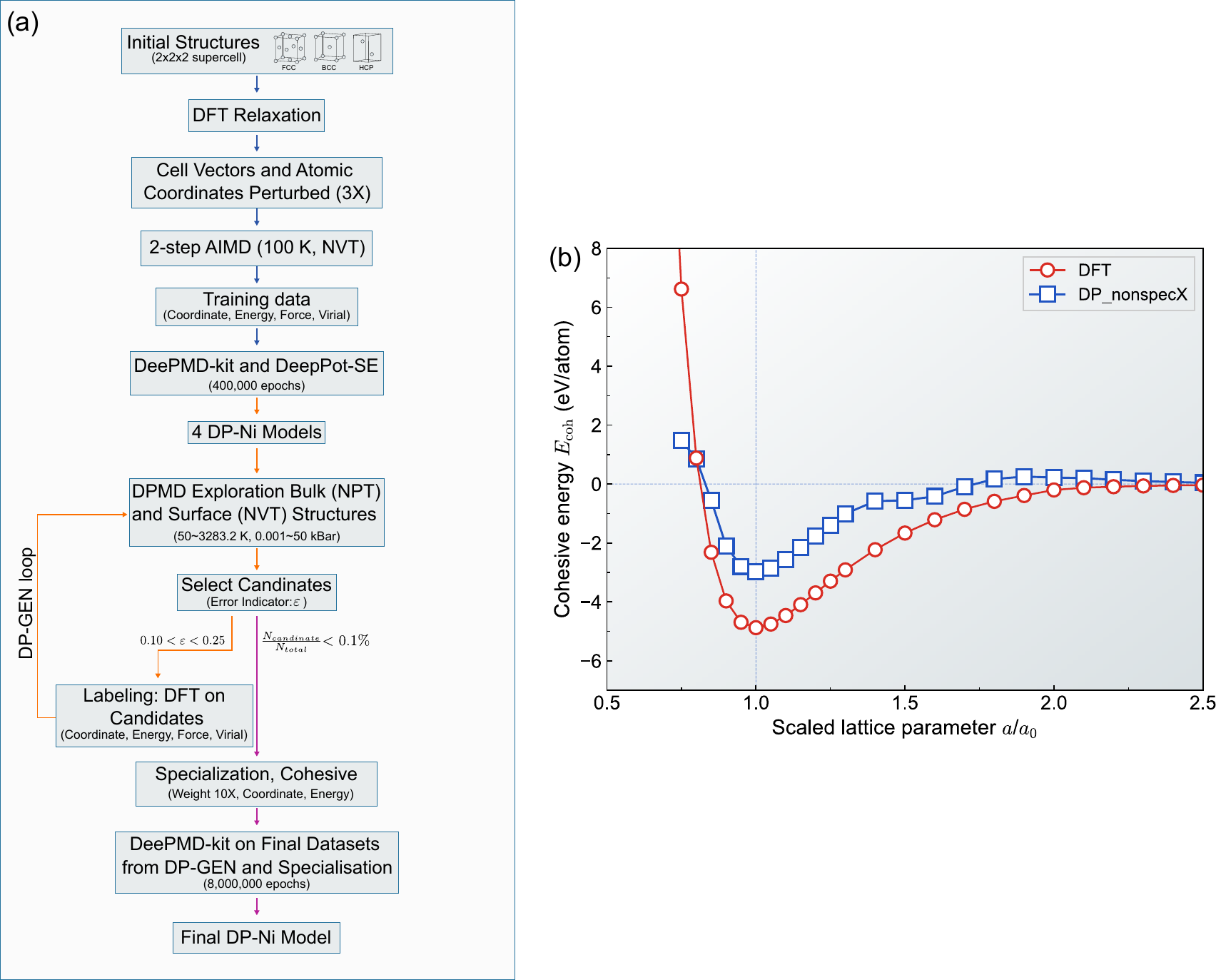}
  \caption{\label{fig:gb} (a) The training workflow of the DP-Ni potential, and (b) the FCC Ni cohesive energy as a function of lattice parameter from DFT and DP-Ni before ``specialization'' (DP\_nonspecX).}
\end{figure*}


  

\end{document}